\documentclass[amsmath,amsfonts,superscriptaddress,prx,preprint,longbibliography]{revtex4-1}
\usepackage{graphicx}
\usepackage{epsfig}
\usepackage{bm}
\usepackage{dcolumn}
\usepackage{amsmath}
\usepackage{amssymb}
\usepackage{amsfonts}
\usepackage{color}
\usepackage{soul}
\usepackage[bookmarks=false,linkcolor=blue,urlcolor=blue,colorlinks,citecolor=blue]{hyperref}

\usepackage{libertine}

\newcommand{\beg}{\begin{equation}}
\newcommand{\en}{\end{equation}}
\newcommand{\bp}{\mathbf p}
\newcommand{\bq}{\mathbf q}

\newcommand{\br}{\mathbf r}

\newcommand \bel  {\begin{align}}
\newcommand \enl  {\end{align}}

\newcommand{\eps}{\epsilon}
\newcommand{\veps}{\varepsilon}

\newcommand{\up}{\uparrow}
\newcommand{\dn}{\downarrow}
\newcommand{\dg}{^\dagger}

\begin{document}

\title{Transport anomalies in multiband superconductors \\ near quantum critical point}

\author{Maxim Dzero}
\affiliation{Department of Physics, Kent State University, Kent, Ohio 44242, USA}

\author{Maxim Khodas}
\affiliation{Racah Institute of Physics, Hebrew University of Jerusalem, Jerusalem 91904, Israel}

\author{Alex Levchenko}
\affiliation{Department of Physics, University of Wisconsin-Madison, Madison, Wisconsin 53706, USA}

\date{November 24, 2023}

\begin{abstract}
We study the effects of quantum fluctuations on the transport properties of multiband superconductors near a pair-breaking quantum critical point. For this purpose, we consider a minimal model of the quantum phase transition in a system with two nested two-dimensional Fermi surfaces. Under the assumption that doping the system adds nonmagnetic impurities but does not change the densities of carriers, we include disorder potentials that render both intra- and interband collisions. Interband scattering leads to full suppression of the unconventional $s^{\pm}$ superconducting order similar to the effect of paramagnetic impurities in isotropic single-band superconductors. We use the diagrammatic technique of quantum field theory to compute the corrections to electrical conductivity in a normal state due to superconducting fluctuations in the entire low-temperature quantum regime. We show that the sign of the conductivity correction depends on how the quantum critical point is approached in the phase diagram. We contrast our findings to existing approaches to this problem based on the renormalization group, time-dependent Ginzburg-Landau phenomenology, and effective bosonic action field theories.    
\end{abstract}

\maketitle

\section{Introduction}

Quantum phase transitions \cite{Sondhi-RMP97,Vojta-RPP2003,Sachdev-QPT-Book} and quantum criticality \cite{Fradkin-Book,Sachdev-QPM-Book} in the strongly-correlated and mesoscopic electron systems represent active areas of research in contemporary condensed matter physics, which are continuously motivated by a multitude of experimental discoveries (see e.g. recent reviews \cite{Matsuda2014,Taillefer-ARCMP2019,KapitulnikRMP19} and references therein). 

In this paper, we address a particular problem of the pair-breaking quantum phase transition in a multiband unconventional superconductor when the transition temperature $T_c(x)$ is suppressed to zero as a function of control parameter $x$ that could be, e.g., chemical doping, externally applied pressure, strain, or magnetic field. The critical value of the control parameter $x_\textrm{c}$ at which $T_c(x_\textrm{c})=0$ defines the superconducting quantum critical point (QCP). We are primarily interested in transport anomalies at the onset of such a quantum phase transition (QPT). The main question that we answer in this paper is: What are the signatures of quantum superconducting fluctuations in dc conductivity when QCP is approached from a normal metal?   

This problem has rich history. It dates back to the pioneering work of Abrikosov and Gor'kov \cite{AG1961} who predicted quantum phase transition in a conventional $s$-wave superconductor contaminated by paramagnetic impurities. It was shown that, provided sufficient concentration of impurities in a system, pair-breaking induced by the spin-flip scattering suppresses critical temperature to zero leading to a gapless state of a superconductor. Even at the level of thermodynamic properties \cite{Galitski-PRL2008}, the complexity of this mechanism is highly nontrivial as it was realized only recently that this transition is of a topological character \cite{Yerin-EPL2022}. 

For the kinetic coefficients at temperatures sufficiently close to critical, $T-T_c\ll T_c$, transport properties of the metallic phase are dominated by superconducting fluctuations \cite{LarkinVarlamovBook}. The principal corrections to the normal-state conductivity are given by the pair-fluctuation Aslamazov-Larkin term \cite{ALOriginal}, interference Maki-Thompson term \cite{Maki-PTP1968,Thompson-PRB1970}, density of states contributions calculated by Abrahams \textit{et al.} \cite{Abrahams-PRB1970}, and diffusion constant renormalization identified by Al'tshuler \textit{et al.} \cite{AVR-JETP1983}. The microscopic origin and physical essence of these terms are well understood. The Aslamazov-Larkin (AL) process corresponds to a new transport channel mediated by the thermally-induced preformed Cooper pairs. The Maki-Thompson (MT) contribution describes the interaction-driven quantum interference of quasiparticles experiencing Andreev reflections on superconducting droplets above $T_c$. The density of states (DOS) mechanism captures the depletion in the single-particle states at the Fermi level once the system is tuned towards $T_c$ thus causing the suppression of conductivity. The final contribution (DCR) can be understood as the renormalization of the single particle diffusion coefficient in the presence of fluctuations. A question of what happens with the interplay of these contributions to the conductivity in the quantum critical regime when $T_c\to0$ constitutes to a genuinely complicated problem. It turns out the the answer depends on how the QCP is approached on the $x-T$ phase diagram defined by the control parameter $x$ and the temperature $T$. In order to properly place our paper in the context of related studies we briefly survey prior results. We will purposefully narrow the scope of this discussion to the two-dimensional (2D) systems while acknowledging that there is equally broad literature devoted to the fluctuational transport in quantum wires (see e.g. \cite{Zaikin,Rosenow,Rogachev-NP18}). 

There are several examples in the literature when fluctuation-driven transport in superconductors was studied in the quantum regime. Perhaps the most considered model by multiple authors \cite{Galitski-PRB2001,Levchenko-PRB2009,GVV-PRB2011,Tikhonov-PRB2012,Michaeli-PRB2012} is that of a disordered thin superconducting film placed in the perpendicular magnetic field. In this case the QCP is realized near the upper critical field $H_{c2}$ when film is driven to the normal state at the lowest temperatures. In the ultra-quantum limit, $T/T_c\ll (H-H_{c2})/H_{c2}$, it was shown in the framework of a diagrammatic perturbation theory that the total fluctuation correction $\delta\sigma$ to the normal-state Drude conductivity $\sigma_D$ is negative in the diffusive limit, and depends on the external magnetic field logarithmically  $\delta\sigma/\sigma_D\propto\ln[(H-H_{c2})/H_{c2}]$ \cite{Galitski-PRB2001}. This example is special since fluctuations occur in the presence of strong Landau quantization. In particular, AL term is less singular than usual, due to large cyclotron gap, and as a result is of the same order as MT and DOS contributions. Alternatively, the film can be driven normal by an in-plane magnetic field. This case splits into two possibilities whether orbital \cite{ShahLopatin-PRL2005,ShahLopatin-PRB2007} or spin \cite{Aleiner-PRL1997,Catelani-PRB2006,KLC-PRL2012} effects limit superconductivity. In the former case, the model is analogous to the Abrikosov-Gor'kov (AG) pair-breaking scenario of QPT, and conductivity turns out be nonsingular near the QCP \cite{ShahLopatin-PRL2005}. In the latter case, large Zeeman gap $E_z$ suppresses fluctuation Cooper pairs exponentially, simply due to a Boltzmann factor $e^{-E_z/T}\ll1$; however, quantum virtual transitions are suppressed only algebraically, $T/E_z\ll1$, and thus contribute to the conductivity correction, which happens to be logarithmic \cite{KLC-PRL2012}. Other examples include fluctuations in the unconventional superconductors with quenched nonmagnetic disorder, which also lead to the QPT since the Anderson theorem \cite{Anderson1959} is not applicable in this case. 

An effective action approach augmented by the renormalization group analysis has led to the conductivity correction $\delta\sigma\propto 1/T$ in the case of a $d$-wave superconductor \cite{Revaz-PRL1997}. This result was in part confirmed based on the quantum extension of the time-dependent Ginzburg-Landau formalism \cite{Mineev-PRB2001}, however, attributed to a high-temperature asymptote of the conductivity correction rather than a quantum critical behavior near the QCP. In contrast, universal dc conductivity was predicted in Ref. \cite{Herbut-PRL2000} at the QCP, $\sigma\simeq e^2/h$. A weakly divergent positive conductivity correction was found in the case of a $p$-wave superconductor \cite{Adachi-JPSJ2001}, $\delta\sigma/\sigma_D\propto \ln[\ln(1/T\tau)]$, where $\tau$ is the elastic scattering time. 
The issue of limits, namely taking the external frequency to zero first followed by taking temperature to zero or vice versa, was discussed in Ref. \cite{Podolsky-PRB2007} where it was shown that the resulting conductivity depends on the order of these limits and particle-hole asymmetry. Finally, we highlight insightful studies \cite{Feigelman,Spivak} on the quantum superconductor-normal-metal phase transition in a system of superconducting grains embedded in a normal metal. These latter studies describe the phase diagram but do not discuss transport phenomena near the superconducting QCP. 

\begin{figure*}[t]
\includegraphics[width=0.85\linewidth]{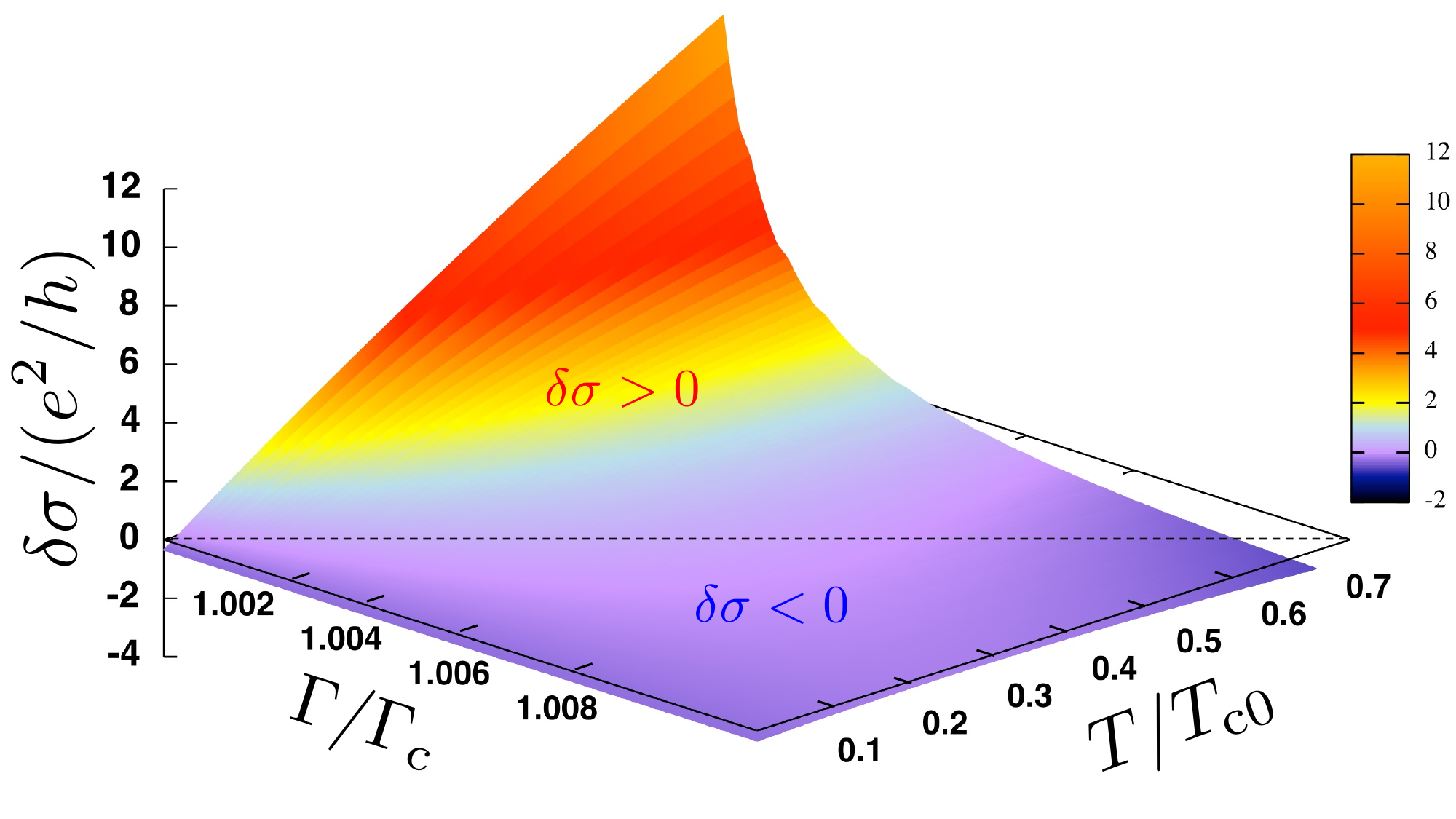}
\caption{The temperature and disorder dependence of the leading quantum correction to conductivity near superconducting quantum critical point.}
\label{Fig-Main1}
\end{figure*}

\begin{figure*}[t]
\includegraphics[width=0.75\linewidth]{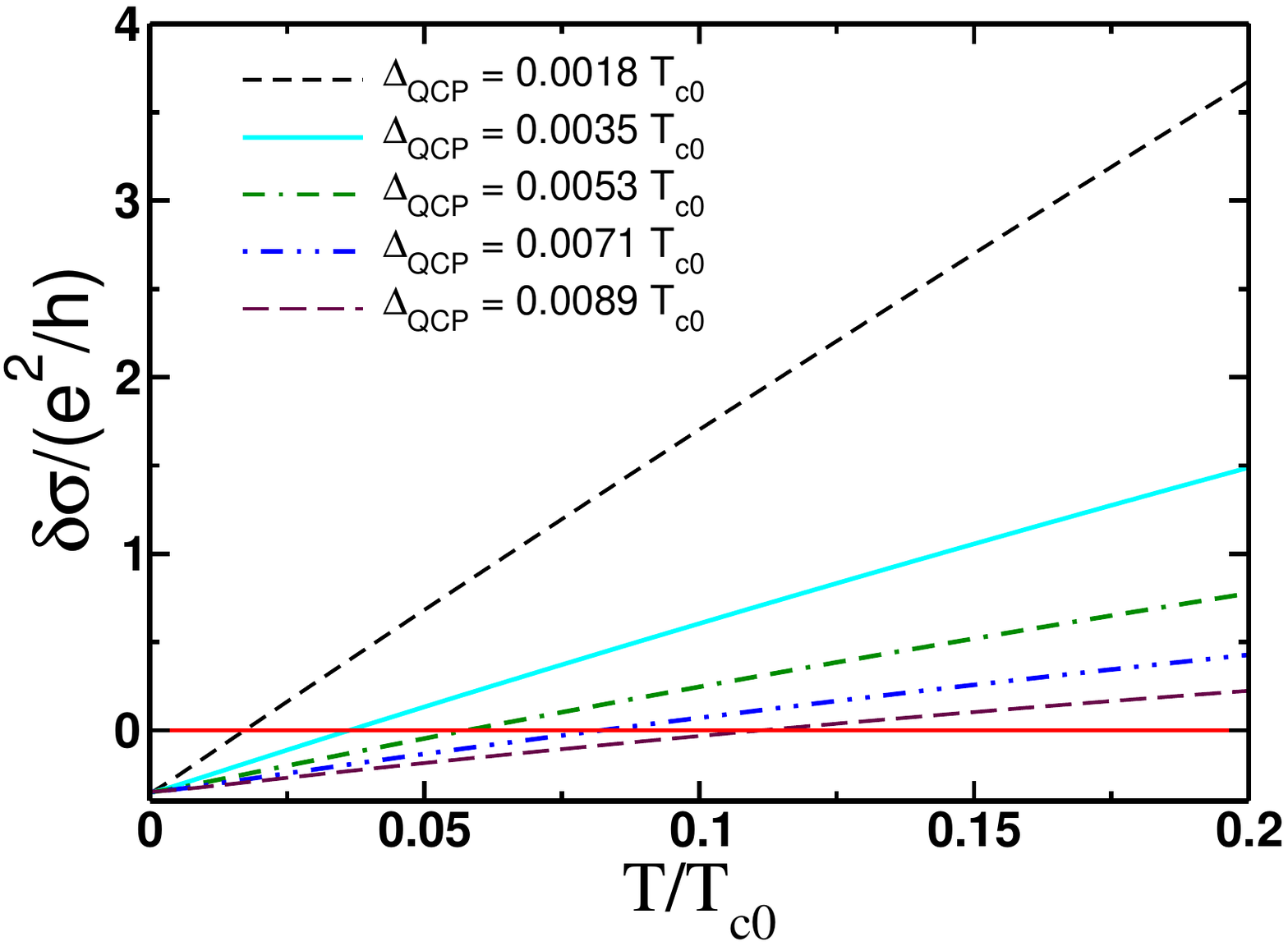}
\caption{The temperature dependence of quantum correction to conductivity for various values of parameter $\Delta_{\textrm{QCP}}\propto \varGamma-\varGamma_c$ describing the system's proximity to a pair-breaking quantum phase transition from a superconducting state. Note that in the quantum regime when $T\ll \Delta_{\textrm{QCP}}$ the quantum correction to conductivity is negative, while in the opposite (thermal) regime it is positive.}
\label{Fig-Main2}
\end{figure*}

In the context of the literature mentioned above, we revisit the problem of quantum transport near the superconducting quantum critical point. 
It should be noted that all the above-mentioned models of the pair-breaking QPT are dual to each other, therefore the apparent lack of consensus in the final result between the different approaches \cite{Revaz-PRL1997,Mineev-PRB2001,Herbut-PRL2000,Podolsky-PRB2007,Adachi-JPSJ2001} is problematic. This gives an additional important motivation to our paper as we attempt to reach a coherent view of a problem. Our approach is based on the direct perturbative diagrammatic technique where we retain both fermionic quasiparticles and bosonic soft modes. We find that existing bosonic effective action theories miss dynamical (frequency dependent) vertex corrections. This omission leads to spurious conclusions. For instance, it is known that even in the regime of thermal fluctuations, bosonic action captures properly only the Aslamazov-Larkin contribution. Generally, the gradient expansion of an effective field theory has to be carefully carried out to describe all the contributions \cite{Levchenko-TDGL} and to the best of our knowledge this has not been demonstrated in the quantum regime thus far. We acknowledge that fluctuation corrections to the transport in a multiband metals were considered in the past \cite{Koshelev-PRB2005,Ramires-JPCM2011} but not in the quantum regime. Furthermore, in the quantum regime, usual approximations to the AL vertex do not apply as one has to keep full account of its frequency dependence. Its analytical structure in the causal retarded-advanced sector is crucial and leads to finite corrections at $T\to0$. In this paper, we show that quantum corrections to conductivity display nonmonotonic temperature dependence. It is dominated by the regular MT, DOS and DCR processes at $T\to0$. The total conductivity correction is negative (localizing) despite the attractive interactions in the Cooper channel. This correction is universal, modulo logarithmic factor, and temperature independent when the temperature is lower than detuning from the QCP in proper units. Curiously, this is qualitatively consistent with the observed saturation of conductivity in the anomalous (strange) metal phases \cite{KapitulnikRMP19}. At slightly higher temperatures, the conductivity correction is positive and dominated by the AL channel of fluctuations. If a system is tuned within the critical fan near QCP and the temperature is lowered, the conductivity is determined by the universal scaling law. However, this regime is beyond the perturbation theory. Our conclusions are limited by a cut-off at exponentially small temperatures as we do not consider localization effects. Figures \ref{Fig-Main1} and  \ref{Fig-Main2} summarize our main finding: they depict the conductivity correction near the QCP in units of conductance quantum plotted as a function of temperature ($T/T_{c0}$) and disorder strength $\varGamma/\varGamma_c$ (units of $T_{c0}$, $\varGamma$ and $\varGamma_c$ on this plot are introduced and explained below in Sec. \ref{Sec:Propagator} see Table \ref{Table-Parameters}).          


The presentation is organized as follows. Inspired by the physics of iron pnictides, in Sec. \ref{sec:Model} we formulate the model of an unconventional multiband superconductor that exhibits magnetic and superconducting QCPs. In Sec. \ref{Sec:Propagator} we derive the two-particle Green's function -- the propagator in the Cooper channel that captures superconducting fluctuations in the thermal and quantum regimes. In Sec. \ref{Sec:Diagrams} we present a direct linear-response computation of the leading corrections to the normal-state conductivity near the QCP based on the Kubo formula and analytical continuation procedure \cite{AGD}. In Sec. \ref{sec:Discussion} we summarize our main findings and provide a broader perspective on a problem extending the discussion to the context of anomalous metal phases observed at the superconductor-to-insulator QPT. Finally, various technical aspects of this paper are presented in Appendices A, B, C, and D. Throughout the paper we use the natural units and set Planck’s and Boltzmann’s constants to unity $k_B=\hbar=1$.

\section{Model of a quantum-critical multiband superconductor}\label{sec:Model}

Motivated by the physics of iron-based superconductors \cite{Chubukov2012,Matsuda2014}, we consider a two-band metallic system in two spatial dimensions (2D). The Hamiltonian we choose to work with can be written as
\beg\label{Eq1}
\hat{H}=\hat{H}_0+\hat{H}_{\textrm{int}}+\hat{H}_{\textrm{dis}}.
\en
The first term represents the noninteracting part of the model
\beg\label{Eq1H0}
\begin{split}
\hat{H}_0&=\sum\limits_{\bp\sigma}\left[\xi_c(\bp)\hat{c}_{\bp\sigma}\dg \hat{c}_{\bp\sigma}+\xi_f(\bp)\hat{f}_{\bp\sigma}\dg\hat{f}_{\bp\sigma}\right].
\end{split}
\en
Here $\hat{c}_{\bp\sigma}(\hat{c}_{\bp\sigma}\dg)$ and $\hat{f}_{\bp\sigma}(\hat{f}_{\bp\sigma}\dg)$ are the single-particle annihilation (creation) operators for each band in a state with momentum $\bp=(p_x,p_y)$ and spin projection $\sigma=(\uparrow,\downarrow)$,  $\xi_c(\bp)=\xi_\bp=\bp^2/2m-\mu$, 
$\xi_f(\bp)=-\xi_\bp$ are the single-particle energies in each of the two corresponding bands relative to the chemical potential $\mu$, $m$ is an effective mass. The parabolic spectrum is chosen here for simplicity of the subsequent calculations. We also neglect the anisotropy in effective masses for the same reason. These approximations do not limit generality of our considerations and final conclusions.   

The second term in Eq. \eqref{Eq1} accounts for the effects of interactions. For simplicity, we take a short-range contact-interaction model described by the Hamiltonian  
\beg\label{Eq1H1}
\begin{split}
\hat{H}_{\textrm{int}}&=\frac{1}{2}\sum\limits_{\alpha\beta\alpha'\beta'}\sum\limits_{\textrm{a,b}=(c,f)}\int V_{\alpha\beta\beta'\alpha'}^{\textrm{ab}}\delta(\br-\br')\overline{\Psi}_{\textrm{a}\alpha}(\br)\overline{\Psi}_{\textrm{a}\beta}(\br)
{\Psi}_{\textrm{b}\beta'}(\br'){\Psi}_{\textrm{b}\alpha'}(\br')d^2\br d^2\br',
\end{split}
\en
where we introduced the four-component spinor with the Fourier components
\beg\label{Spinor}
\overline{\Psi}_{\bm{k}}=(\hat{c}_{\bm{k}\uparrow}\dg, \hat{c}_{-\bm{k}\downarrow}\dg,
\hat{f}_{\bm{k}\uparrow}\dg,\hat{f}_{-\bm{k}\downarrow}\dg).
\en
The matrix elements of the interaction $V_{\alpha\beta\beta'\alpha'}^{\textrm{ab}}$ are defined according to
\beg\label{Interact}
\begin{split}
V_{\alpha\beta\beta'\alpha'}^{(\textrm{ab})}=g_{\textrm{ab}}\left(i\hat{\sigma}^y\right)_{\alpha\beta}\left[\left(i\hat{\sigma}^y\right)\dg\right]_{\beta'\alpha'},
\end{split}
\en
where $\hat{\sigma}^y$ is the second Pauli matrix and $g_{\textrm{ab}}=\left(1-\delta_{\textrm{ab}}\right)g$.
The sign of $g$ determines whether the superconducting order parameter is $s$ wave or $s^{\pm}$ wave. In what follows, we assume that $g>0$, which corresponds to the extended $s^{\pm}$ symmetry of the superconducting order parameter. Here we obviously ignore the other interaction channels such as ones that may lead to the spin- or charge-density-wave instabilities. This approximation is sufficient to capture the QCP at the end point of the superconducting dome, which is far from the region of parameters in the phase diagram where magnetic interactions play an important role.  

The third term in Eq. \eqref{Eq1} accounts for the disorder scattering with an assumption of point-like impurity potential:
\beg\label{Hdis}
\begin{split}
\hat{H}_{\textrm{dis}}&=
\sum\limits_{i\textrm{a}}\int d^2\br\left\{u_0\overline{\Psi}_{\textrm{a}\alpha}(\br)\Psi_{\textrm{a}\beta}(\br)+u_\pi\overline{\Psi}_{\textrm{a}\alpha}(\br)\Psi_{\overline{\textrm{a}}\beta}(\br)\right\}\hat{\sigma}_{\alpha\beta}^z\delta(\br-{\mathbf R}_i).
\end{split}
\en
The convention $\overline{c}=f$, $\overline{f}=c$ is implied, while ${\mathbf R}_i$ label coordinates of randomly distributed scatterers. In this expression $\hat{\sigma}^z$ is the third Pauli matrix. The first term in Eq. \eqref{Hdis} $(\propto u_0)$ describes the intraband impurity scattering, while the second term $(\propto u_\pi)$ accounts for the interband impurity scattering. For simplicity, we neglect possible anisotropy of impurity potentials as it only leads to the redefinition of transport scattering times.    

\begin{figure}[t!]
  \centering
 \includegraphics[width=0.75\linewidth]{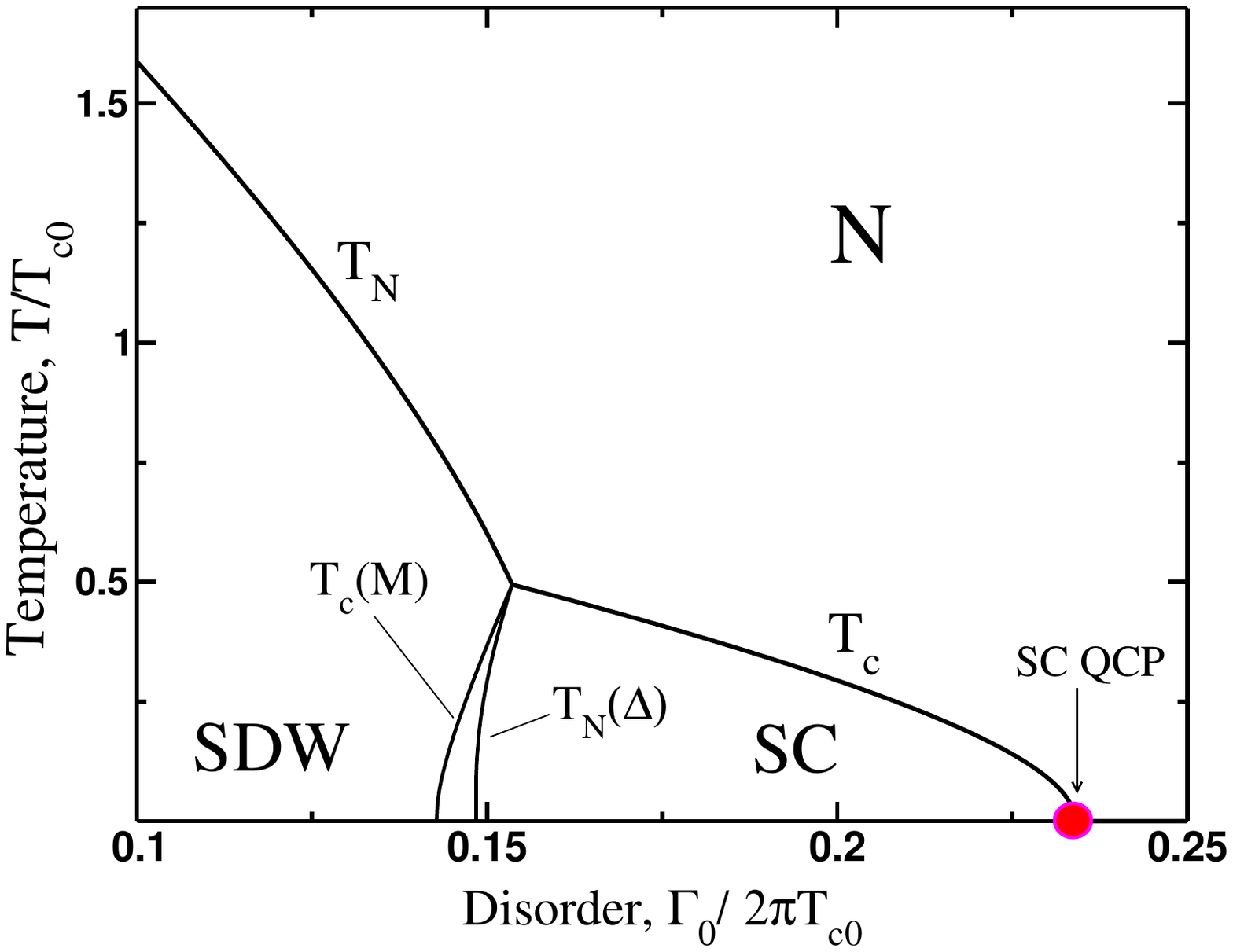}
  \caption{A representative phase diagram which can be obtained from the mean-field analysis of the two-band model formulated in Sec. \ref{sec:Model} with an addition of the interactions, which induce the spin-density-wave instability. Here $T_{\textrm{c0}}$ is the critical temperature of a superconducting transition in a clean system, $T_N$ is the N\'eel temperature of the spin-density-wave (SDW) transition, $T_c$ is the superconducting critical temperature in a disordered system, $\Delta$ is the superconducting order parameter, $M$ is a staggered magnetization of the SDW order, $\varGamma_0=2\pi\nu_F|u_0|^2$ is the intraband disorder scattering rate, Eq. (\ref{Hdis}), and $\nu_F$ is the single-particle density-of-states at the Fermi level. The end point of the superconducting dome at the high doping marks the pair-breaking QCP, which is labeled by a red dot on a plot.}
  \label{Fig1}
\end{figure}

A few comments are in order about the model formulated in this section. With an addition of the interaction, which drives the spin-density-wave instability, the model defined by Eq. (\ref{Eq1}) can be used to reproduce a set of quite generic experimental observations in several compounds belonging to the family of the iron-based superconductors \cite{Chubukov2012,Matsuda2014,FS-PRB2010}. The mean-field analysis of this model \cite{VC-PRB2011,Fernandes-PRB2012,VVC} shows that there is always a region of the phase diagram where superconductivity coexists with the spin-density-wave order, see Fig. \ref{Fig1} and Refs. \cite{GrinenkoNP2020,Carvalho-PRB2020}. The model also allows one to incorporate the quantum fluctuation effects near the spin-density-wave transition (the one which is at the border between the coexistence region and purely superconducting phase), which are responsible for the nonmonotonic dependence of the London penetration depth on the amount of chemical substitutions \cite{Hashimoto,Auslaender,Joshi,ChoSciAdv2016,ZhengPRL2018,Levchenko2013,Chowdhury2013}. Specifically, it has been demonstrated that quantum fluctuations produce linear-in-temperature dependence of the London penetration depth with a slope that becomes steeper as the system is tuned toward the quantum critical point (QCP) \cite{Khodas-PRB2020,Hasan-PRB2022}, as well as the power-law dependence in the heat capacity \cite{Carvalho-PRB2020} all on the background of the exponential temperature dependence found at the mean-field approximation. The effect of quantum fluctuations on kinetic coefficients of iron pnictides near quantum critical points is less explored theoretically, while this topic is highly motivated experimentally, see e.g., Refs. \cite{Hayes-NP16,Maksimovic-PRX20,Malinowski:2020,Hayes-NP21}. In part to fill this gap, below we will follow the avenues of Refs. \cite{Khodas-PRB2020,ShahLopatin-PRL2005,ShahLopatin-PRB2007,KLC-PRL2012,GVV-PRB2011} and discuss the effects of system's proximity to the end point of superconducting dome (see Fig. \ref{Fig1}) in transport using the model Hamiltonian of Eq. (\ref{Eq1}). 


\section{Pairing fluctuation propagator}\label{Sec:Propagator}

The central quantity in our analysis will be the propagator for the superconducting fluctuations. First we define the single-particle correlation function in the imaginary time formalism \cite{AGD}
\beg\label{G12}
\begin{split}
{\cal G}_{\textrm{a}\sigma}(x,x')=-\left\langle\hat{T}_\tau\left\{\Psi_{\textrm{a}\sigma}(x)\overline{\Psi}_{\textrm{a}\sigma}(x')\right\}\right\rangle
\end{split}
\en
with $x=(\br,\tau)$. In the Matsubara frequency representation the expressions for the Green's functions, which also include the effects of disorder are given by \cite{Dzero2015,Dzero2021} 
\beg\label{Gcf}
{\cal G}_{\textrm{c(f)}}(\bp,i\varepsilon_n)=\frac{1}{i\varepsilon_n\mp\xi_\bp},\qquad
\varepsilon_n=\epsilon_n+\frac{1}{2\tau_{\textrm{t}}}\textrm{sign}\left(\epsilon_n\right),
\en
where 
\begin{equation}\label{eq:Gamma-t}
\varGamma_{\text{t}}\equiv\tau_{\textrm{t}}^{-1}=2\pi\nu_F\left(|u_0|^2+|u_\pi|^2\right)\equiv\tau_0^{-1}+\tau_\pi^{-1}
\end{equation} is the total scattering rate, and we have omitted the spin index for brevity.

The expression for the superconducting fluctuations propagator $\hat{L}$ can be formally obtained by introducing the two-particle correlation function and developing the perturbative expansion in powers of the coupling constant $g_{\textrm{ab}}$ \cite{AGD,LarkinVarlamovBook}. The resummation of the ladder diagrammatic series yields the following expression
\beg\label{Lab}
\hat{L}^{-1}(\bq,i\omega_l)=\hat{\Pi}(\bq,i\omega_l)-\hat{V}^{-1},
\en
where $\hat{\Pi}(\bq,i\omega_l)$ is the response function to the external pairing perturbation 
$\chi_{\textrm{a}}\Psi_{\textrm{a}\up}\dg \Psi_{\textrm{a}\dn}\dg+\textrm{H.c.}$ in each of the two bands, $\bq=(q_x,q_y)$, $\omega_l=2\pi Tl$ $(l=0,\pm 1,\pm 2,...)$ are bosonic Matsubara frequencies, $T$ is temperature. The inverse interaction matrix we choose in the form
\beg\label{Vmat}
\hat{V}^{-1}=\frac{\nu_F}{\lambda}\left(\begin{matrix} 0 & 1 \\ 1 & 0\end{matrix}\right)=\frac{\nu_F}{\lambda}\hat{\sigma}^x,
\en
where $\lambda=\nu_Fg$ is the dimensionless coupling constant corresponding to the pairing in the $s^{\pm}$ pairing channel. For brevity, we take equal density of states and equal pairing constants in each band. This simplification does not limit generality of our results. For the diagonal and off-diagonal elements of the matrix function $\hat{\Pi}(\bq,i\omega_l)$ we found the following expressions:
\beg\label{Pab}
\begin{split}
\Pi_{\textrm{aa}}(\bq,\omega_l)-\Pi_{\textrm{a}\overline{\textrm{a}}}(\bq,\omega_l)\approx \nu_F\left[\ln\left(\frac{2\gamma_E\omega_D}{\pi T}\right)-\varPsi(\bq,\omega_l)\right], \qquad 
\Pi_{\textrm{a}\overline{\textrm{a}}}(\bq,\omega_l)\approx{\nu_F}\varXi(\bq,\omega_l),
\end{split}
\en
where $\ln\gamma_E\approx 0.577$ is the Euler constant and $\omega_D$ is the ultraviolet cut-off of the theory, which can be absorbed to define $T_c$ (see below). Additionally, we have introduced the following auxiliary functions
\begin{subequations}\label{betas}
\begin{align}
&\varPsi(\bq,\omega_l)=\psi\left(\frac{1}{2}+\frac{1}{2\pi T\tau_\pi}+\frac{|\omega_l|+{\cal D}\bq^2}{4\pi T}\right)-\psi\left(\frac{1}{2}\right), \\
&\varXi(\bq,\omega_l)=\frac{1}{2}\left[\psi\left(\frac{1}{2}+\frac{1}{2\pi T\tau_\pi}+\frac{|\omega_l|+{\cal D}\bq^2}{4\pi T}\right)-\psi\left(\frac{1}{2}+\frac{|\omega_l|+{\cal D}\bq^2}{4\pi T}\right)\right].
\end{align}
\end{subequations}
Here $\psi(z)$ is the logarithmic derivative of the Euler's gamma function and ${\cal D}=v^2_F\tau_{\text{t}}/2$ is the diffusion coefficient. In the following it will be convenient to change the basis using the unitary transformation
\beg\label{Utrans}
\hat{U}=\hat{U}^{-1}=\frac{1}{\sqrt{2}}\left(\begin{matrix} 1 & \phantom{+}1 \\ 1 & -1\end{matrix}\right).
\en
Thus, the expression for the fluctuation propagator in this new basis is  
\beg\label{NewProp}
\hat{L}^{-1}(\bq,\omega_l)=\left(\begin{matrix} \Pi_{\textrm{aa}}+\Pi_{\textrm{a}\overline{\textrm{a}}} +\frac{\nu_F}{\lambda} & 0 \\ 0 & \Pi_{\textrm{aa}}-\Pi_{\textrm{a}\overline{\textrm{a}}}-\frac{\nu_F}{\lambda}
\end{matrix}\right).
\en
Matrix element $(\hat{L}^{-1})_{11}$ accounts for the fluctuations in the $s$-wave channel, while the remaining matrix element describes the fluctuations in the $s^{\pm}$-wave channel. Indeed, if we choose $\lambda<0$ then the critical temperature is independent of the interband scattering rate. We can interpret this as a manifestation of the Anderson theorem \cite{Anderson1959,AG1961}.

Inverting Eq. (\ref{NewProp}) and performing the unitary transformation yields 
\beg\label{hatLfinfin}
\begin{split}
\hat{L}(\bq,\omega_l)&=\frac{\nu_F^{-1}\hat{\sigma}^0}{{\cal W}(\bq,\omega_l)}+
\frac{\nu_F^{-1}\left[\varPsi(\bq,\omega_l)+\frac{1}{\lambda}\right](\hat{\sigma}^x-\hat{\sigma}^0)}{{\cal W}(\bq,\omega_l)\left[\ln({T}/{T_{c0}})+\varPsi(\bq,\omega_l)\right]},
\end{split}
\en
where $T_{c0}$ is the critical temperature of the superconducting transition in a clean superconductor and ${\cal W}=\nu_F^{-1}\Pi_{\textrm{aa}}+\varXi+1/\lambda$.
Note that the first term on the right-hand side of Eq. (\ref{hatLfinfin}) is nonsingular and therefore will be ignored in what follows. Furthermore, in the weak-coupling limit we can also approximate
\beg\label{ApproxPreFactor}
\frac{\varPsi(\bq,\omega_l)+\frac{1}{\lambda}}{{\cal W}(\bq,\omega_l)}\approx 1.
\en
Thus, we arrive to the following approximate expression for the matrix function, which describes the pairing fluctuations in the two-band model Eq. (\ref{Eq1}):
\beg\label{ApproxL}
\hat{L}(\bq,\omega_l)\approx\frac{\nu_F^{-1}}{\ln({T}/{T_{c0}})+\varPsi(\bq,\omega_l)}(\hat{\sigma}^x-\hat{\sigma}^0).
\en
In passing we note that the same expression can be obtained from the semiclassical Usadel equations in the limit of strong disorder when $T\tau_{\textrm{t}}\ll 1$. 

Expressions above allow one to compute the critical temperature of the superconducting transition $T_c$ as well as the critical value of the pair-breaking rate $\tau_{\pi}^{(c)}$ at a given temperature by setting $\ln({T}/{T_{c0}})+\varPsi(\bq=0,i\omega_l=0)=0$. We find that $T_c\to 0$ when $\tau_\pi$ reaches the critical value
\beg\label{Tc0}
\left[\tau^{(c0)}_\pi\right]^{-1}\approx\pi T_{\textrm{c0}}/2\gamma_E.
\en
In order to find a finite-temperature correction to $\tau_\pi^{(c0)}$, we assume that $T\tau_{\pi}^{(c0)}\ll 1$ and expand the digamma functions
using the asymptotic expression $\psi(z)\approx\ln z-1/2z-1/(12z^2)$, $(z=1/2\pi T\tau_\pi)$. A rather straightforward calculation yields
\beg\label{taupicFin}
\frac{\tau_\pi^{(c)}}{\tau_{\pi}^{(c0)}}\approx 1-\frac{\pi^2}{6}\left(T\tau_{\pi}^{(c0)}\right)^2.
\en
The ``distance" from/to the critical pair-breaking point can now be defined according to
\beg\label{deltapb}
\delta_{\textrm{QCP}}(T)=\left[\frac{\tau_\pi^{(c)}(T)}{\tau_\pi}-1\right]\equiv{\Delta_{\textrm{QCP}}(T)}\tau_{\pi}^{(c)}(T).
\en
Whether the quantum critical fluctuations or thermal fluctuations have a dominant effect on transport will be determined by the relation between 
parameter $|\delta_{\textrm{QCP}}(T)|$ and another dimensionless parameter $T\tau_\pi^{(c)}(T)$. We also introduce the critical scattering rate $\varGamma_c=1/\tau^{(c)}_\pi$. These parameters are summarized in Table \ref{Table-Parameters} for convenience and will be helpful in the analysis of the different transport regimes. 

\begin{table}[h!]
\centering
\begin{tabular}{ |c|c|c|} 
 \hline
 Notation & Definition & Equation  \\ 
 \hline\hline
 $\varGamma_0$ & Intraband scattering rate  & $\varGamma_0=2\pi\nu_F|u_0|^2$  \\ 
 \hline
 $\varGamma_\pi$ & Interband scattering rate & $\varGamma_\pi=2\pi\nu_F|u_\pi|^2$  \\ 
 \hline
 $\varGamma_{\text{t}}$ & Total scattering rate & Eq. \eqref{eq:Gamma-t}  \\ 
 \hline
 $\varGamma_c$ & Critical scattering rate & $\varGamma_c=1/\tau^{(c)}_\pi$  \\ 
 \hline
 $T_{c0}$ & Transition temperature & $T_{c0}\simeq\omega_De^{-1/\lambda}$  \\ 
 \hline
 $\Delta_{\text{QCP}}$ & Energy gap to the QCP & Eq. \eqref{deltapb}  \\ 
 \hline
 $\mathcal{D}$ & Diffusion coefficient & $\mathcal{D}=v^2_F\tau_{\text{t}}/2$  \\ 
 \hline
 $\sigma_{D}$ & Drude conductivity& $\sigma_D=2e^2\nu_F\mathcal{D}$  \\ 
 \hline
\end{tabular}
  \caption{Summary of key parameters of the model. The notation of $\varGamma_i$ defines a particular scattering rate distinguished by a subscript 
 $i=0,\pi,\text{t}, c$ whose meaning is given in the column of definitions. Each scattering rate can be equivalently rewritten in terms of the corresponding scattering time, namely $\varGamma_i=\tau^{-1}_i$.}
 \label{Table-Parameters}
\end{table}

\begin{figure}[t]
  \centering
 \includegraphics[width=0.75\linewidth]{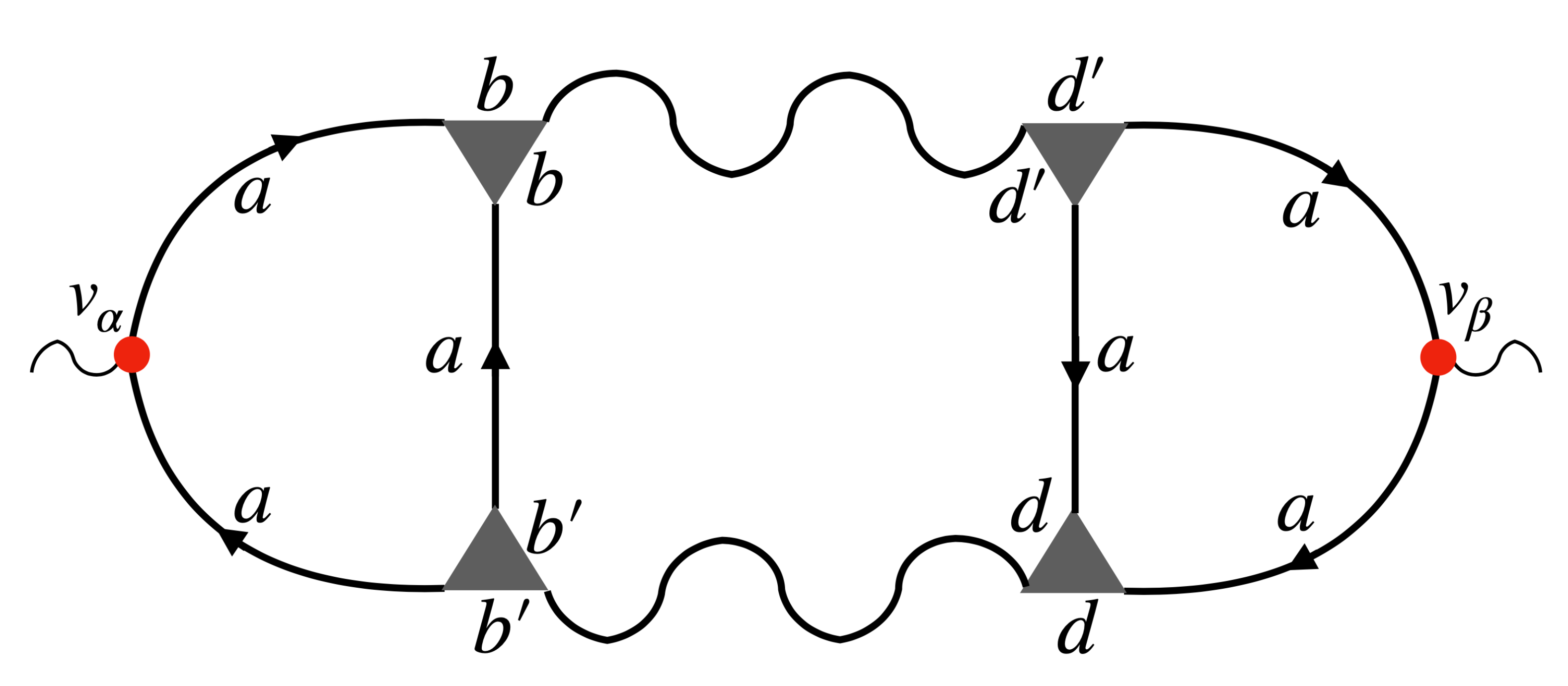}
  \caption{The Aslamazov-Larkin diagram for the correction to conductivity from the fluctuation-driven Cooper pairs. The Latin letters designate the band indices. Solid lines are single-particle propagators, wavy lines are pairing fluctuation propagators, shaded triangles are Cooperon vertex functions and solid circles are the velocity operators.}
  \label{Fig2AL}
\end{figure}


\section{Fluctuation corrections to the conductivity near QCP}\label{Sec:Diagrams}

In this section we present a step-by-step derivation of all the superconducting fluctuation corrections to the conductivity at the one-loop level. We first derive general expressions and then analyze their limiting cases. We demonstrate that our two-band model is analogous to the QPT a realized in the superconducting thin film subjected to the in-plane magnetic field with the orbital effect of pair breaking. When appropriate, we rectify previously made statements about the temperature dependence of various contributions to conductivity that emerge in this theory, and highlight terms 
that were overlooked in the previous computations. 

\subsection{Aslamazov-Larkin correction}
We begin by considering the following correlation function in the Matsubara representation:
\beg\label{xKorr}
K_{\alpha\beta}(\omega_n)=-\int\limits_0^{1/T}d\tau e^{i\omega_n\tau}\langle\hat{T}_\tau\hat{J}_\alpha(\tau)\hat{J}_\beta(0)\rangle,
\en
where $\hat{J}_\alpha$ is the $\alpha=x,y$ component of the current operator. In this definition brackets denote both quantum statistical and disorder average, and $\hat{T}_\tau$ defines the ordering in imaginary time. The $\textrm{dc}$ part of conductivity is then determined by analytically continuing the response function 
$K_{\alpha\beta}(\omega_n)$ to real frequencies using $\omega_n\to -i\omega$,
\beg\label{Kubo}
\sigma_{\alpha\beta}=\lim\limits_{\omega\to 0}\left[\frac{K_{\alpha\beta}(-i\omega)}{-i\omega}\right].
\en
The analytical expression for the Aslamazov-Larkin diagram (Fig. \ref{Fig2AL}) takes the form  
\beg\label{KAL}
K_{\alpha\beta}^{\textrm{AL}}(\omega_n)=-4e^2T\sum\limits_{\omega_l}\sum\limits_{\textrm{abb}'\textrm{dd}'}\int\frac{d^2\bq}{(2\pi)^2}
\Lambda_{\textrm{abb}'}^{(\alpha)}(\bq;\omega_l,\omega_n)
\Lambda_{\textrm{ad}\textrm{d}'}^{(\beta)}(\bq;\omega_l,\omega_n)
L_{\textrm{bd}'}(\bq,\omega_l)L_{\textrm{b}'\textrm{d}}(\bq,\omega_l-\omega_n).
\en
In this expression functions $\Lambda_{\textrm{abb}'}^{(\alpha)}(\bq;\omega_l,\omega_n)$ represent the triangular blocks each of which consists of three single-particle propagators and two Cooperons,
\beg\label{Lamab}
\Lambda_{\textrm{abb}'}^{(\alpha)}(\bq;\omega_l,\omega_n)=T\sum\limits_{\veps_m}\Gamma_{\textrm{a}}^{(\alpha)}(\bq;\omega_l,\omega_n,\veps_m)
C_{\textrm{ab}}(\bq;\omega_l-\veps_m,\veps_m)C_{\textrm{ab}'}(\bq;\omega_l-\veps_m,\veps_m-\omega_n),
\en
where 
\beg\label{Gamab}
\Gamma_{\textrm{a}}^{(\alpha)}(\bq;\Omega_l,\omega_n,\veps_m)=
\int\frac{d^2\bp}{(2\pi)^2}v_{\alpha}^{(\textrm{a})}(\bp){\cal G}_{\textrm{a}}(\bp,\veps_m){\cal G}_{\textrm{a}}(\bp,\omega_n+\veps_m){\cal G}_{\textrm{a}}(\bq-\bp,\omega_l-\veps_m),
\en
and $v_{\alpha}^{(\textrm{a})}(\bp)$ is the single-particle velocity in each band. In the band-basis vertex functions (Cooperons) $C_{\textrm{ab}}({\mathbf q},\veps_1,\veps_2)$ naturally have diagonal and off-diagonal components. In the basis in which matrix $\hat{C}$ is diagonal, its components describe Cooperons in the $s$-wave and $s^{\pm}$-wave channels correspondingly. It will be convenient to define the components of $\hat{C}(\bq;\veps_1,\veps_2)$ as 
\begin{subequations}\label{Coops}
\begin{align}
&{C}_{\textrm{aa}}+{C}_{\textrm{a}\overline{\textrm{a}}}=\frac{\tau_{\textrm{t}}^{-1}\theta(-\veps_1\veps_2)}{|\veps_1-\veps_2|+{\cal D} \bq^2}, \\ 
&{C}_{\textrm{aa}}-{C}_{\textrm{a}\overline{\textrm{a}}}=
\frac{\tau_{\textrm{t}}^{-1}\theta(-\veps_1\veps_2)}{\frac{2}{\tau_\pi}+|\veps_1-\veps_2|+{\cal D} \bq^2}.
\end{align}
\end{subequations}

Calculation of the triangular block functions $\mathbf{\Lambda}_{\textrm{abb}'}(\bq;\omega_l,\omega_n)$ can be done using the diffusive limit, $T\tau_{\textrm{t}}\ll 1$, with the following result:
\beg\label{LambdaApprox}
\mathbf{\Lambda}_{\textrm{abc}}\approx-4\pi\nu_F{\bq}\left({\cal D}\tau_{\textrm{t}}^2\right)T\sum\limits_{\veps_m}
C_{\textrm{ab}}(\bq;\omega_l-\veps_m,\veps_m)C_{\textrm{ac}}(\bq;\omega_l-\veps_m,\veps_m-\omega_n).
\en
It is clear that due to the symmetry properties of the Cooperons ${C}_{\textrm{aa}}={C}_{\overline{\textrm{a}}\overline{\textrm{a}}}$ and 
${C}_{\textrm{a}\overline{\textrm{a}}}={C}_{\overline{\textrm{a}}\textrm{a}}$, only four out of eight components of $\mathbf{\Lambda}_{\textrm{abc}}$ are independent. 
Because of this property, prior to performing summations over the Matsubara frequencies in Eq. (\ref{KAL}), we sum over the band indices first. Taking
into account Eqs. (\ref{ApproxL}) and (\ref{Coops}), we immediately see that the following significant simplification occurs (we omit the arguments of the corresponding functions for brevity):
\beg\label{GreatSimplify}
\sum\limits_{\textrm{abb}'\textrm{dd}'}\Lambda_{\textrm{abb}'}^{(\alpha)}L_{\textrm{bd}'}L_{\textrm{b}'\textrm{d}}\Lambda_{\textrm{ad}\textrm{d}'}^{(\beta)}=(8\nu_F{\cal D})^2q_\alpha q_\beta L(\bq,\omega_l)L(\bq,\omega_l-\omega_n){\cal K}^2(\bq;\omega_l,\omega_n)
\en
with function ${\cal K}(\bq;\omega_l,\omega_n)$ defined by 
\beg\label{DefK}
\begin{split}
{\cal K}(\bq;\omega_l,\omega_n)=\frac{1}{4\pi\omega_n}&\left\{B(\bq;\omega_l+\omega_n,\omega_n)+B(\bq;\omega_l,\omega_n)\right\}, 
\end{split}
\en
where 
\begin{equation}\label{eq:def-B}
B(\bq;\veps_m,\omega_n)=\psi[(1+x_\bq)/2+(|\veps_m|+\omega_n)/4\pi T]-\psi[(1+x_\bq)/2+|\veps_m|/4\pi T], 
\end{equation}
and 
\beg\label{Defxq}
x_\bq=\frac{{\cal D}\bq^2}{2\pi T}+\frac{1}{\pi T\tau_\pi}\equiv\frac{\varGamma_\bq}{2\pi T}.
\en
We note that function $L(\bq,\omega_l)$ appearing in Eq. (\ref{GreatSimplify}) is given by the prefactor in front of the matrix in Eq. (\ref{ApproxL}). Surprisingly, we find that this expression matches exactly the corresponding expression found for a single-band superconductor in the presence of the paramagnetic impurities in AG model \cite{LarkinVarlamovBook} and so only pair-breaking part of the Cooperon [the second expression in Eq. (\ref{Coops})] contributes to the kernel.

The remaining steps in the calculation are fairly straightforward. As a result, we find four different contributions to conductivity, so for the Aslamazov-Larkin correction we can write
\beg\label{ALfour}
\delta\sigma^{\textrm{AL}}=\delta\sigma_1^{\textrm{AL}}+\delta\sigma_2^{\textrm{AL}}+\delta\sigma_3^{\textrm{AL}}+\delta\sigma_4^{\textrm{AL}}.
\en
The first two terms originate from performing an analytic continuation and expanding pair fluctuation propagator in powers of $\omega$, while the remaining two appear as a result of the expansion of functions $\{B^{RR}(\bq,\varepsilon,\omega), B^{RA}(\bq,\varepsilon,\omega), B^{AA}(\bq,\varepsilon,\omega)\}$ with superscripts denoting retarded ($R$) and advanced ($A$) parts with respect to the energy arguments in Eq. \eqref{eq:def-B}. The first one is a standard contribution and it is given by
\beg\label{sigma1p}
\delta\sigma_1^{\textrm{AL}}=\left(\frac{e\nu_F{\cal D}}{4\pi T}\right)^2\int\frac{q^2d^2\bq}{(2\pi)^2}\int\limits_{-\infty}^{+\infty} \frac{d\omega/2\pi T}{\sinh^2\frac{\omega}{2T}}
\left(\textrm{Re}\left[B'(\bq,\omega)\right]\right)^2 \left[\textrm{Im}L^{R}(\bq,\omega)\right]^2
\en
with $B(\bq,\omega)=\psi[1/2+(\varGamma_\bq-i\omega)/4\pi T]$. The function $L^{R}(\bq,\omega)$ is obtained from $L(\bq,\omega_l)$ by performing an analytic continuation to the real frequencies $\omega_l\to -i\omega$. The expression in Eq. (\ref{sigma1p}) is the so-called thermal contribution, which is governed by the thermal fluctuations in the Cooper channel. 
The second one of the first two contributions to conductivity is 
\beg\label{sigma1pp}
\begin{split}
\delta\sigma_2^{\textrm{AL}}=&\left(\frac{e\nu_F{\cal D}}{4\pi T}\right)^2\!\!\!\int\frac{q^2d^2\bq}{(2\pi)^2}\int\limits_{-\infty}^{+\infty}\frac{d\omega}{\pi}\coth\frac{\omega}{2T}\\&\times
\textrm{Re}\left[\left(\left[B'(\bq,\omega)\right]^2-\left(\textrm{Re}\left[B'(\bq,\omega)\right]\right)^2\right)\frac{\partial}{\partial\omega}\left[L^{R}(\bq,\omega)\right]^2\right].
\end{split}
\en
This correction to conductivity accounts for the contribution from the quantum critical fluctuations, which was missed, along with $\delta\sigma_3^{\textrm{AL}}$ and $\delta\sigma_4^{\textrm{AL}}$, in the earlier studies \cite{Revaz-PRL1997,Mineev-PRB2001,Herbut-PRL2000,Podolsky-PRB2007,Adachi-JPSJ2001}.
\begin{figure*}[t]
\includegraphics[width=0.475\linewidth]{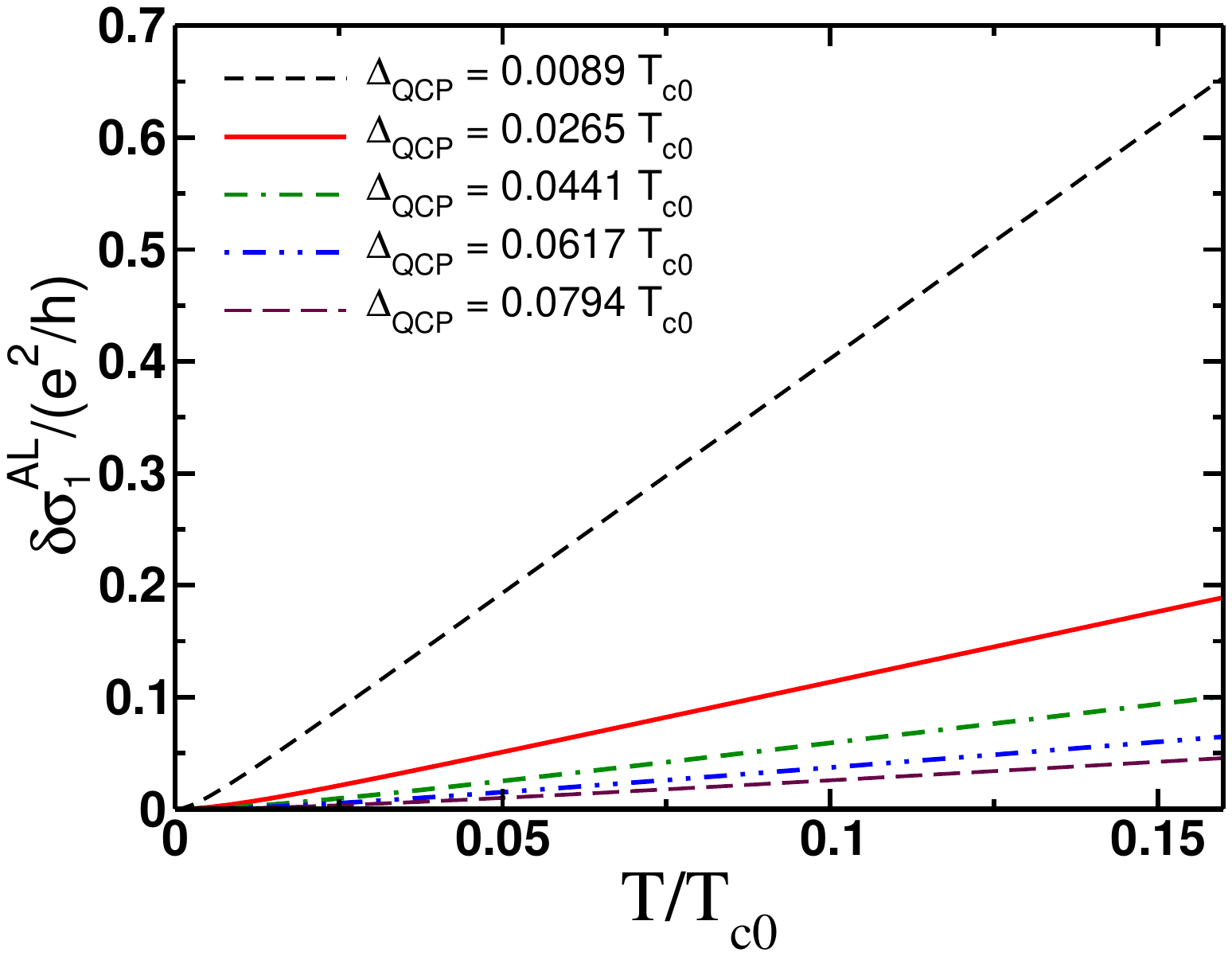}
\includegraphics[width=0.475\linewidth]{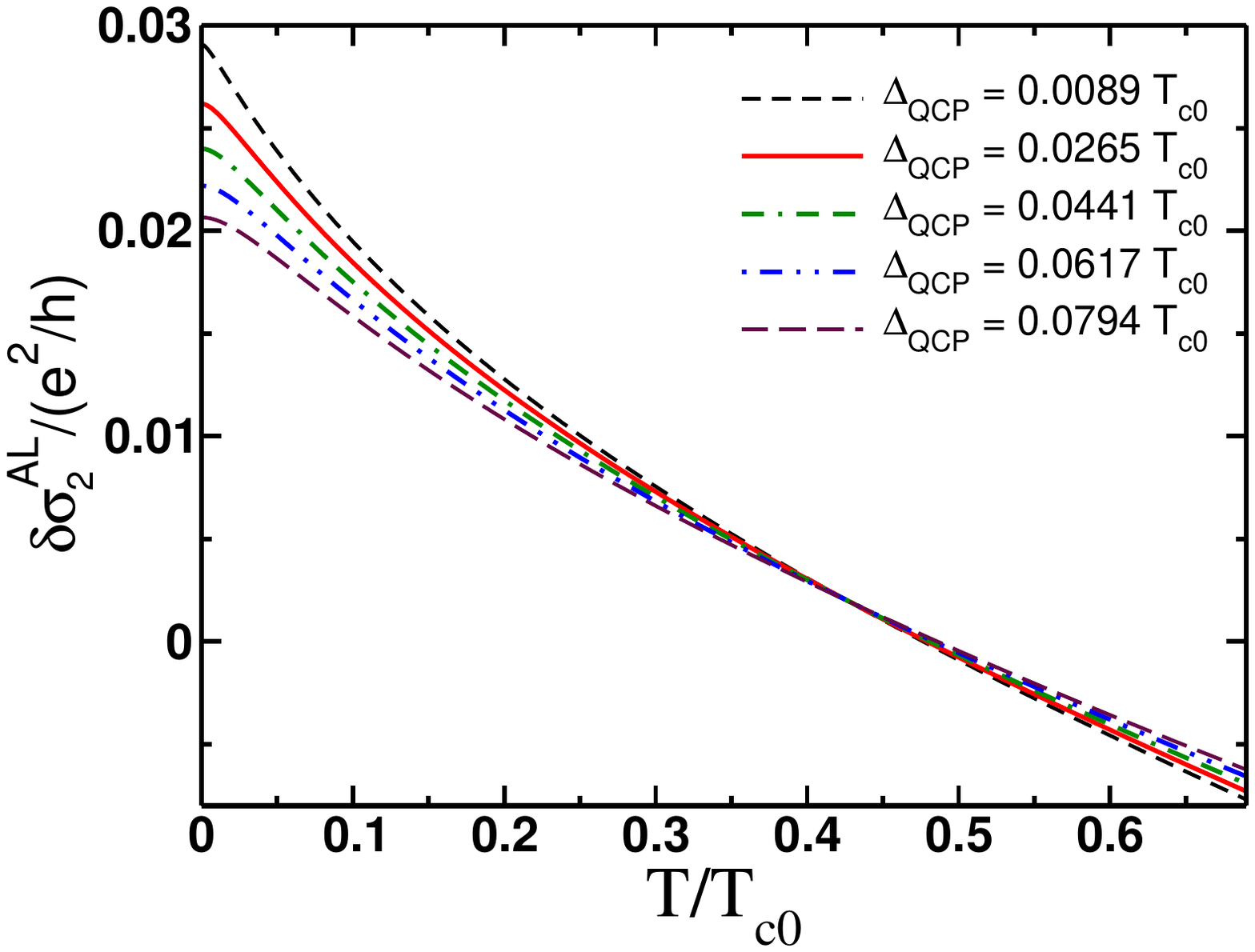}
\includegraphics[width=0.475\linewidth]{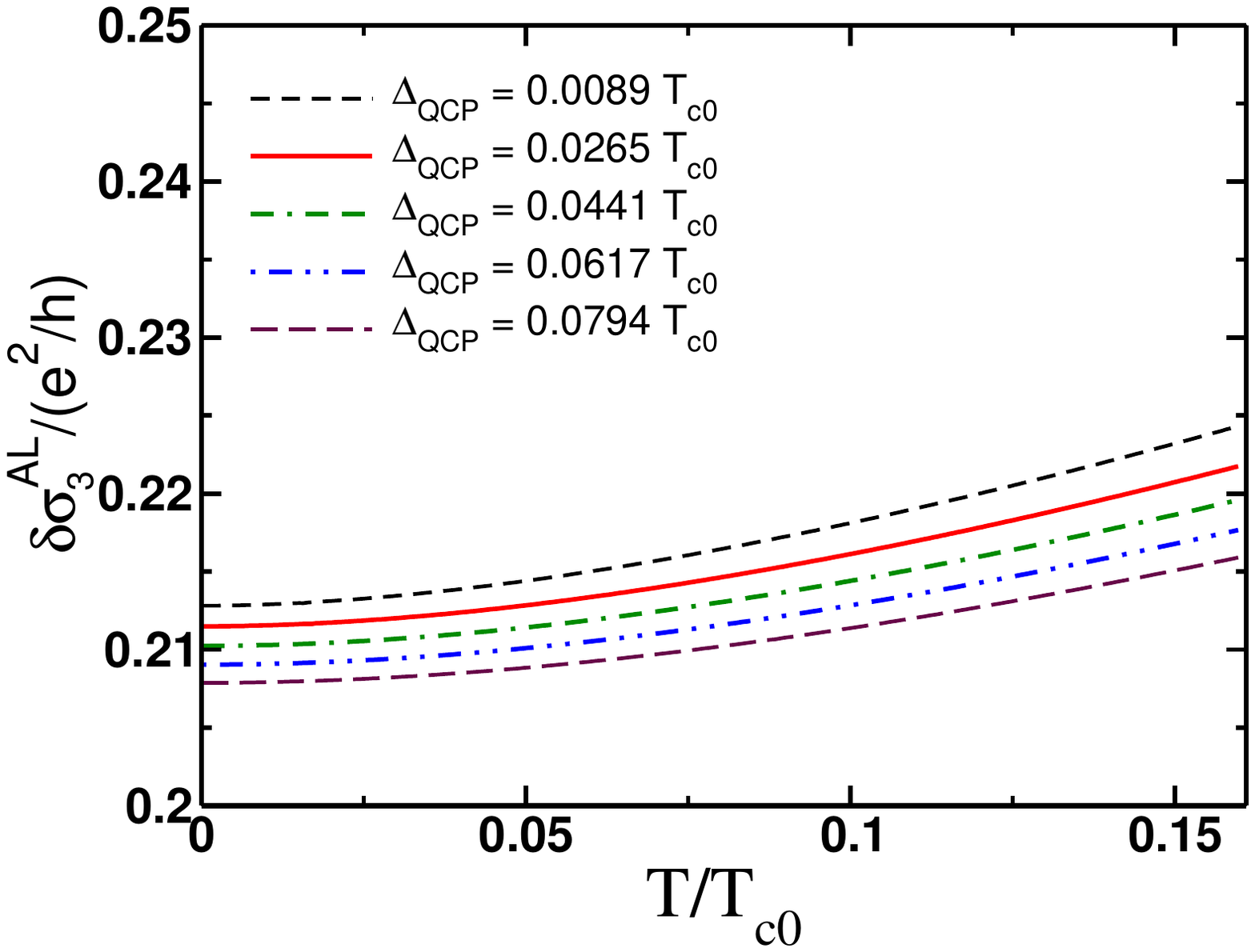}
\includegraphics[width=0.475\linewidth]{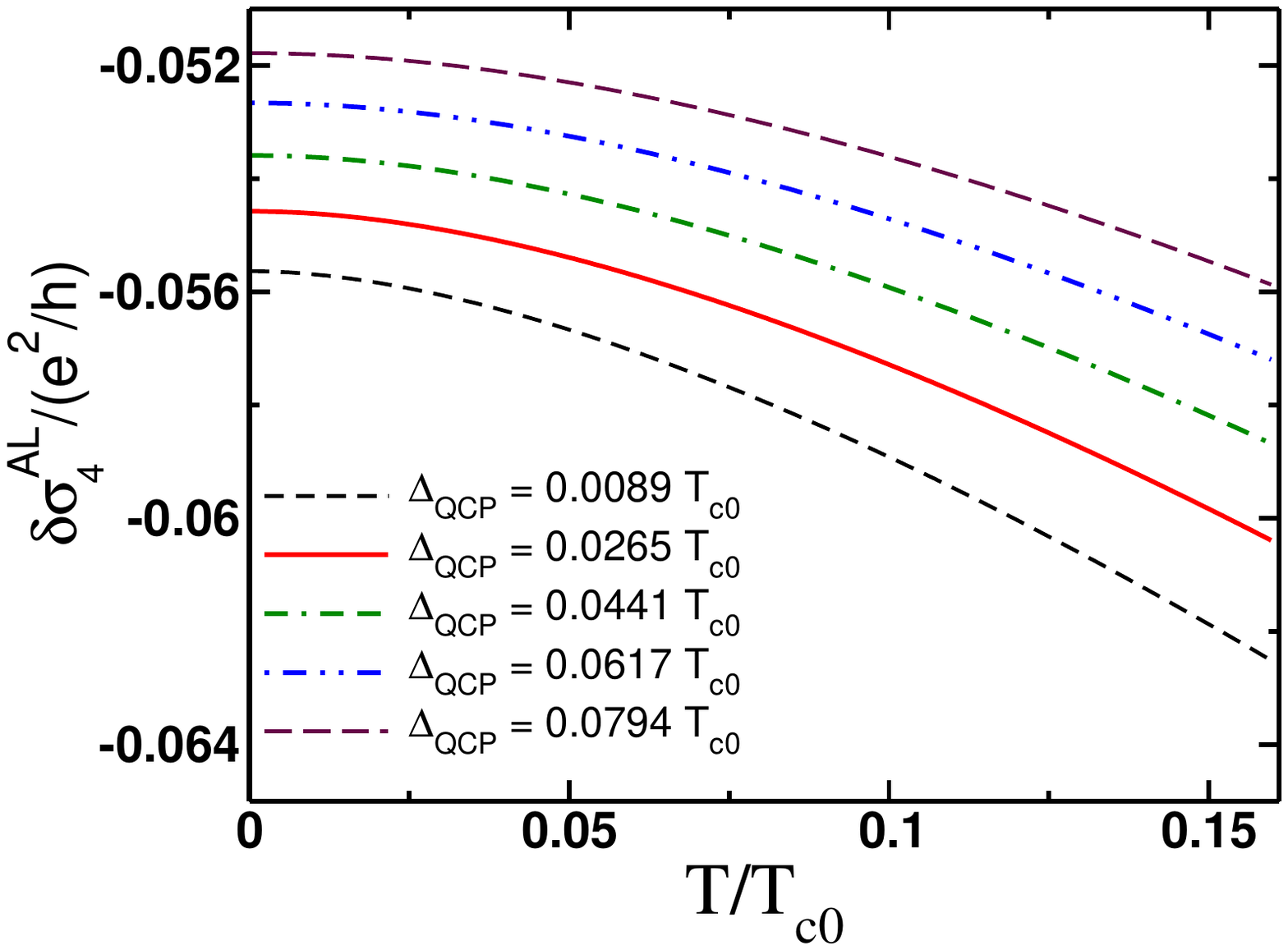}
\caption{Temperature dependence of the four contributions to the quantum correction to conductivity described by the Aslamazov-Larkin diagram, Fig. \ref{Fig2AL}.}
\label{Fig-AL1234}
\end{figure*}
The remaining two contributions originate from the expansion of the functions $\mathbf{\Lambda}_{\textrm{abc}}$ in powers of external frequency $(-i\omega)$. For the first one of the two we find
\beg\label{sigma2p}
\delta\sigma_3^{\textrm{AL}}=-\left(\frac{e\nu_F{\cal D}}{4\pi T}\right)^2\int\frac{q^2d^2\bq}{(2\pi)^2}\int\limits_{-\infty}^{+\infty}\frac{d\omega}{\pi^2T}\coth\frac{\omega}{2T}
\textrm{Im}\left[B'(\bq,\omega)B''(\bq,\omega)\left(L^{R}(\bq,\omega)\right)^2 \right].
\en
Lastly, for the remaining contribution we obtain
\beg\label{sigma2pp}
\delta\sigma_4^{\textrm{AL}}=\left(\frac{e\nu_F{\cal D}}{4\pi T}\right)^2\int\frac{q^2d^2\bq}{(2\pi)^2}\int\limits_{-\infty}^{+\infty}\frac{d\omega}{2\pi^2T}\coth\frac{\omega}{2T}
\textrm{Re}\left[B'(\bq,\omega)\right]\textrm{Im}\left[B''(\bq,\omega)\right]L^{R}(\bq,\omega)L^{A}(\bq,\omega).
\en
These expressions are fairly general and apply in a broad range of parameters that define the superconducting phase diagram (Fig. \ref{Fig1}). The temperature dependence of each of these four terms computed numerically is shown in Fig. \ref{Fig-AL1234}.

\begin{table}[h!]
\centering
\begin{tabular}{ |c|c|c|} 
 \hline
 $\delta\sigma^{\text{AL}}/T$ & $T\lesssim\Delta_{\text{QCP}}\ll T_{c0}$ & $\Delta_{\text{QCP}}\lesssim T\ll T_{c0}$  \\ 
 \hline\hline
 $\delta\sigma^{\text{AL}}_1$ & $\frac{e^2}{6\pi}\frac{T^2}{\Delta^2_{\text{QCP}}}$ & $\frac{e^2}{8\pi}\frac{T}{\Delta_{\text{QCP}}}$  \\ 
 \hline
 $\delta\sigma^{\text{AL}}_2$ & $\frac{33\pi^2-316}{64\pi^2}e^2$ & $-\frac{3e^2}{8\pi}(12\ln2-\frac{49}{6})\left(\frac{T}{\varGamma_c}\right)$  \\ 
 \hline
 $\delta\sigma^{\text{AL}}_3$ & $\frac{e^2}{2\pi^2}$ & $e^2\frac{T}{\varGamma_c}\ln\frac{\varGamma_c}{\Delta_{\text{QCP}}}$  \\ 
 \hline
 $\delta\sigma^{\text{AL}}_4$ & $-\frac{5e^2}{18\pi^2}$ & $-\frac{e^2}{3\pi}\frac{T}{\varGamma_c}$  \\ 
 \hline
\end{tabular}
  \caption{Asymptotic expressions for the leading temperature dependence of the AL contributions in a crossover from quantum-to-thermal regimes of fluctuations as extracted from Eqs. \eqref{sigma1p}--\eqref{sigma2pp}. The temperature dependence of the sum of these four contributions is shown in Fig. \ref{Fig-AL-Total}.}
 \label{Table-AL}
\end{table}

\begin{figure*}[t]
\includegraphics[width=0.75\linewidth]{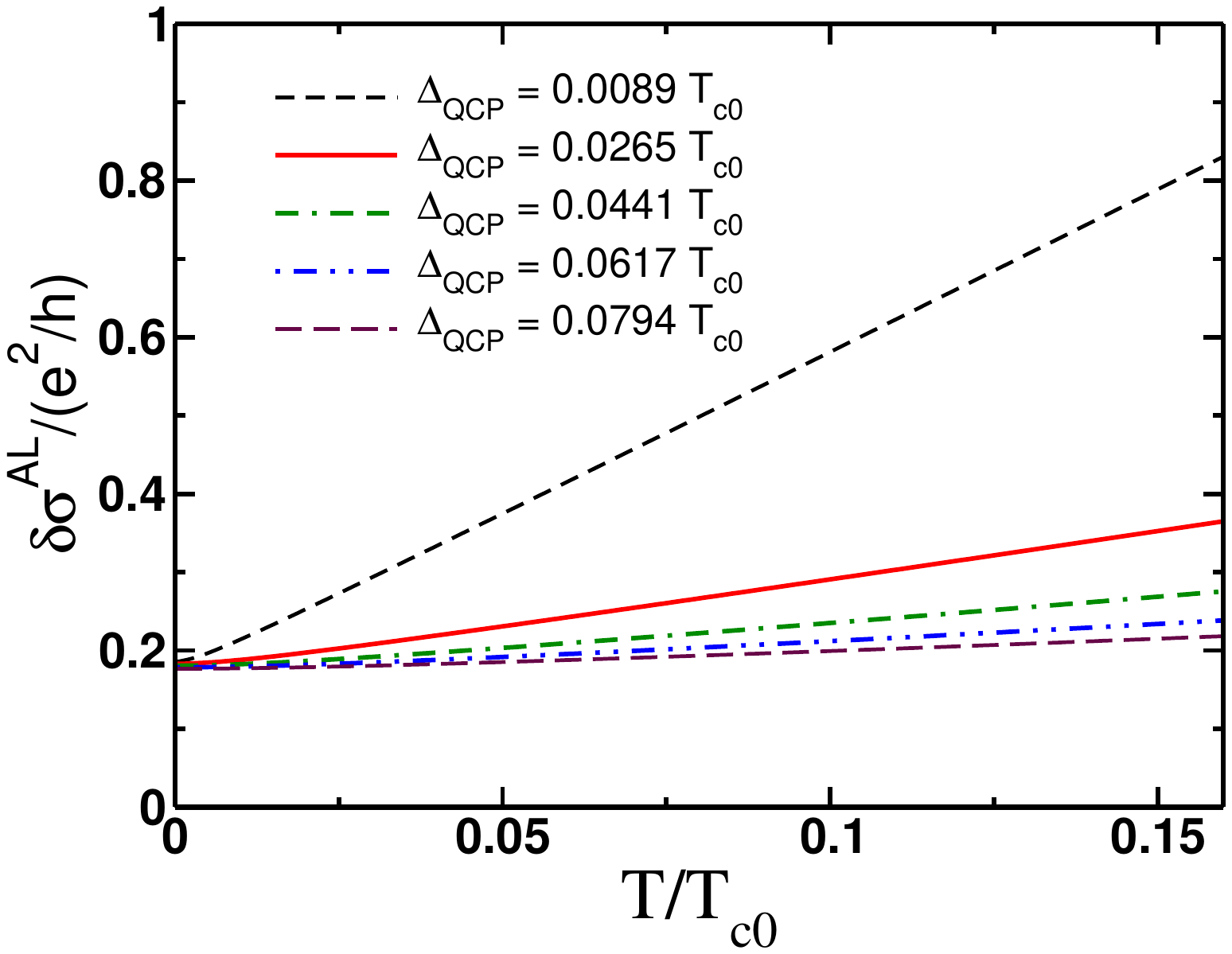}
\caption{Temperature dependence of the total quantum correction to conductivity described by the Aslamazov-Larkin diagram, Fig. \ref{Fig2AL}.}
\label{Fig-AL-Total}
\end{figure*}

We are primarily interested in the low-temperature limit $T\ll T_{c0}$. In this case we still have to distinguish two distinct regimes. We define the \textit{quantum regime} when temperature is smaller than detuning from the quantum critical point, namely $T\ll \Delta_{\text{QCP}}\ll T_{c0}$. We define the intermediate \textit{thermal regime}, when thermal broadening is bigger than the gap to QCP, namely $\Delta_{\text{QCP}}\ll T\ll T_{c0}$. The main results of the calculation for all the AL terms are summarized in Table \ref{Table-AL}, whereas technical details of the derivation are delegated to Appendix \ref{App-AL}. It is noteworthy to observe that except for $\delta\sigma^{\text{AL}}_1$ all other terms remain finite at $T\to0$ limit and overall correction is positive of the order of quantum of conductance, $\delta\sigma^{\text{AL}}(T\to0)\sim e^2$, Fig. \ref{Fig-AL-Total}. As we demonstrate below, the Aslamazov-Larkin contribution is a leading one in the thermal regime $T>\Delta_{\textrm{QCP}}$.

\subsection{Maki-Thompson correction}

\begin{figure}[t]
  \centering
 \includegraphics[width=2.5in]{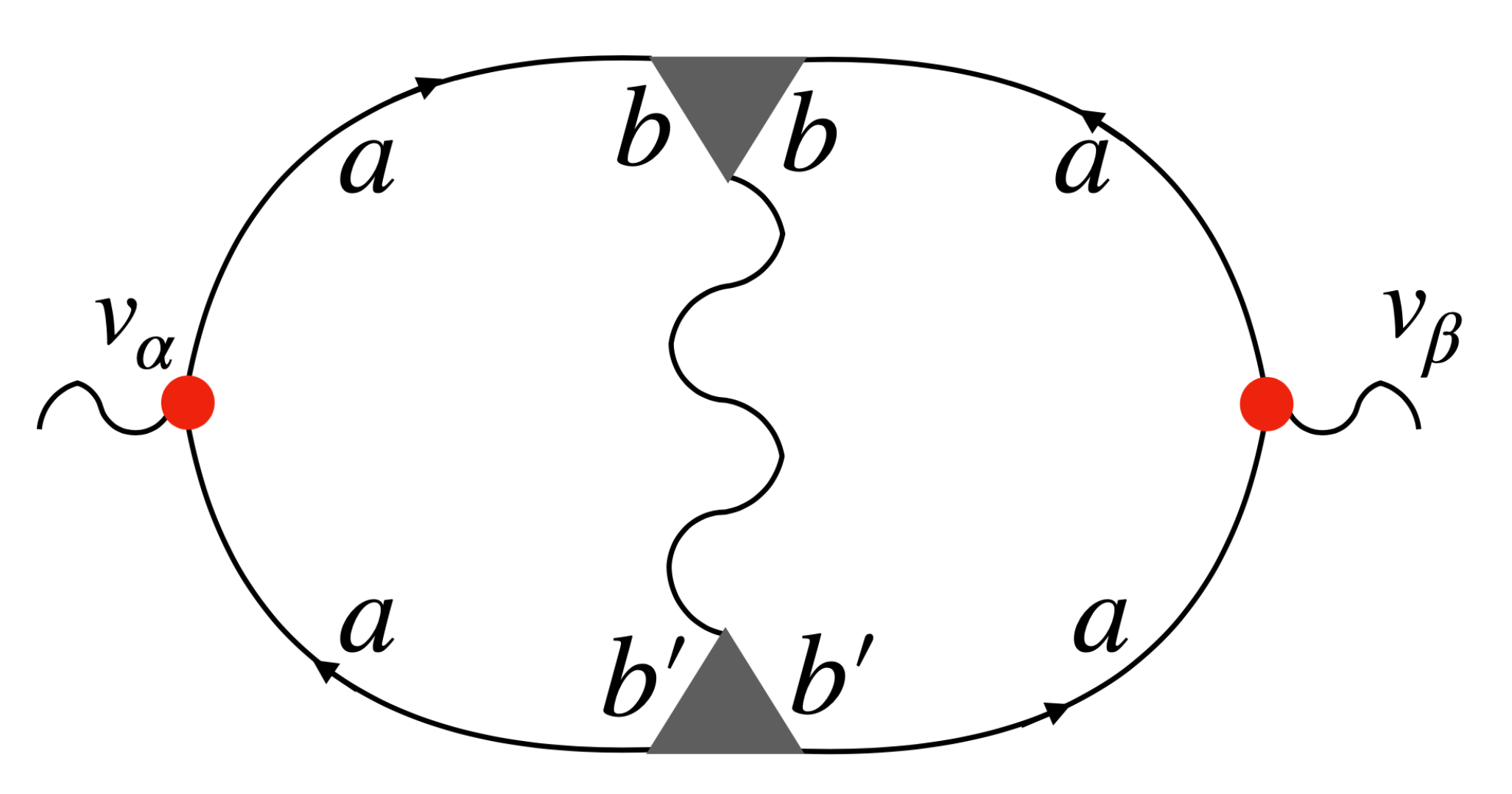}
  \caption{The Maki-Thompson diagram for the fluctuation correction to conductivity at temperatures above the critical temperature. The Latin letters designate the band indices. Solid lines are single-particle propagators, wavy lines are pairing fluctuation propagators, shaded triangles are Cooperon vertex functions, and solid triangles are the velocity operators. Note that we ignore the renormalization of the velocity vertices by disorder since, as it is well known, it leads to the renormalization of the transport time.}
  \label{Fig3MT}
\end{figure}

\begin{figure*}[t]
\includegraphics[width=0.325\linewidth]{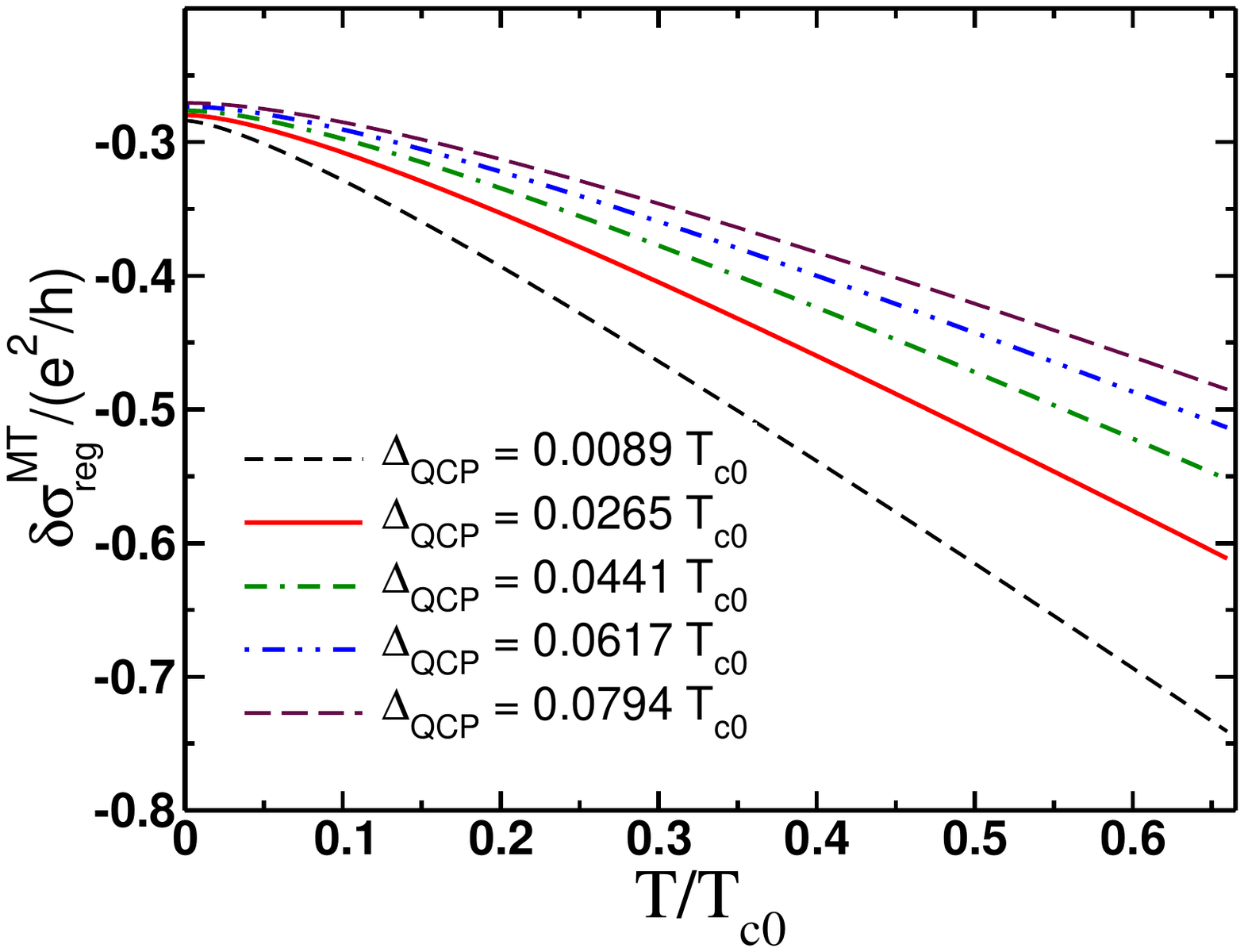}
\includegraphics[width=0.325\linewidth]{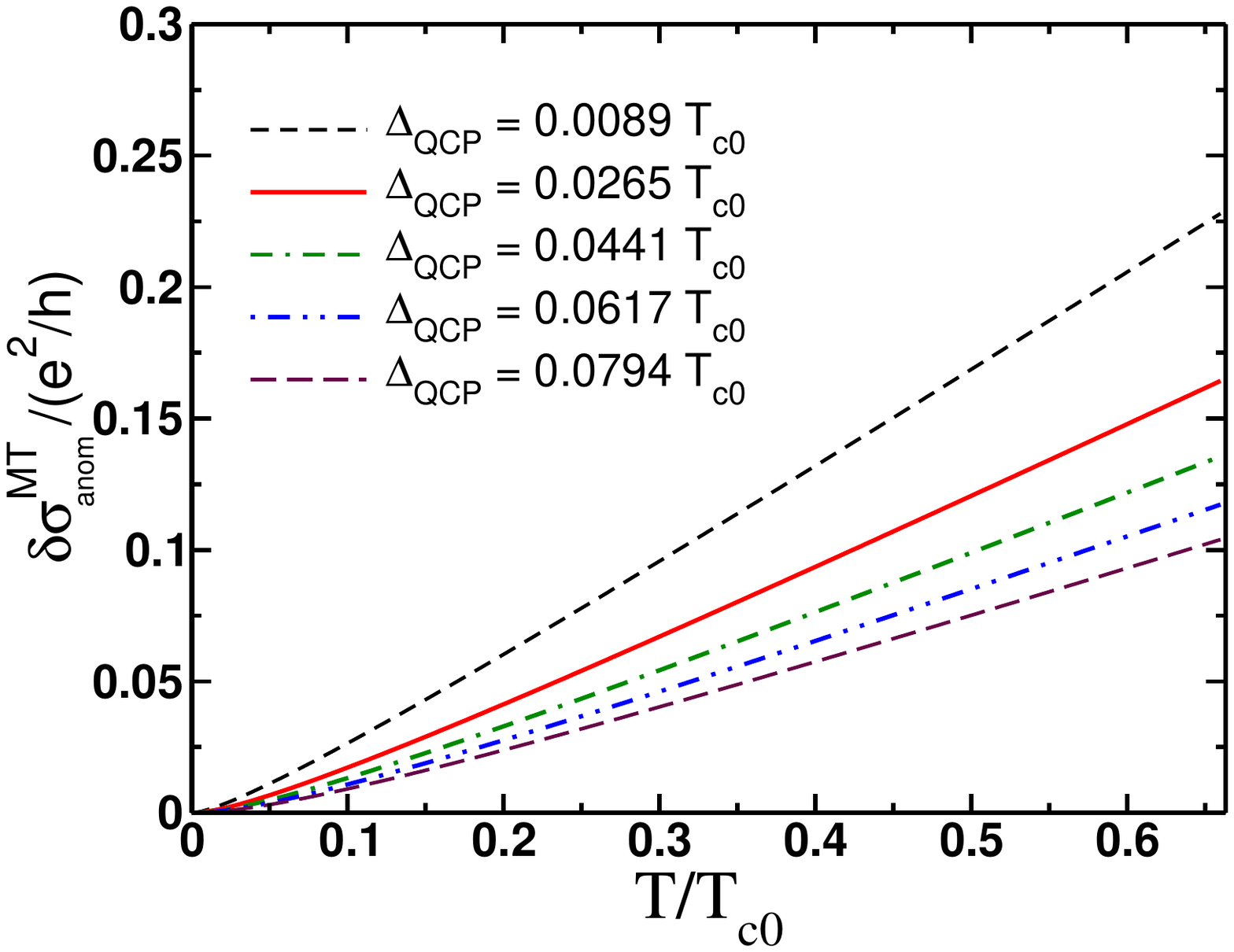}
\includegraphics[width=0.327\linewidth]{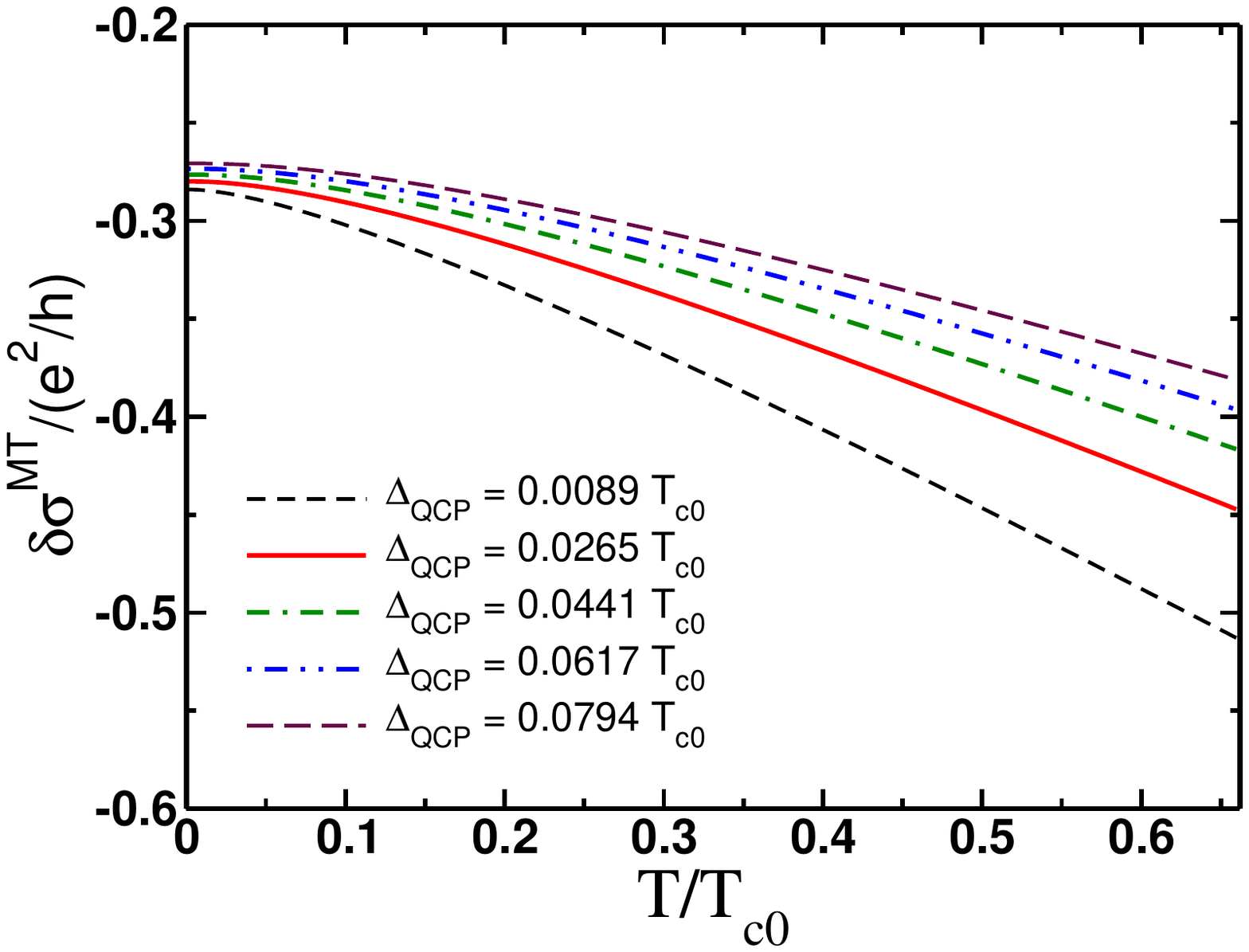}
\caption{\small The temperature dependence of the fluctuation correction to conductivity from the Maki-Thompson diagram, Fig. \ref{Fig3MT}. The first two panels display the contributions from the regular and anomalous terms correspondingly. The third panel shows the total Maki-Thompson contribution. Comparing these results with the results of the analytic calculations shown in Table \ref{Table-MT} we note that the crossover temperature $T^*$ separating quantum and thermal regimes is approximately $T^*\approx 0.15 T_{c0}$.}
\label{Fig-ResMT}
\end{figure*}

We now turn our attention to the diagram, Fig. \ref{Fig3MT}, which describes another pair fluctuation correction to conductivity, the Maki-Thompson contribution. 
As it can be readily verified, the simplification similar to the one Eq. (\ref{GreatSimplify}) occurs in this case as well. For the response function in this case we obtain
\beg\label{KMT}
K_{\alpha\beta}^{\text{MT}}(i\omega_n)=2e^2T\sum\limits_{\omega_l}\int\frac{d^2\bq}{(2\pi)^2}\Sigma_{\alpha\beta}^{\text{MT}}(\bq;i\omega_l,i\omega_n)L(\bq,i\omega_l).
\en
Here function $\Sigma_{\alpha\beta}^{\text{MT}}(\bq;i\omega_l,i\omega_n)$ is determined by the following expression:
\beg\label{FirstMT}
\begin{split}
&\Sigma_{\alpha\beta}^{\text{MT}}(\bq;i\omega_l,i\omega_n)=T\sum\limits_{\eps_k}\lambda(\bq;\eps_{k+n},\omega_l-\eps_{k+n})\lambda(\bq;\eps_k,\omega_l-\eps_k)\\&\times\int\frac{d^2\bp}{(2\pi)^2}v_\alpha(\bp)v_\beta(\bq+\bp){\cal G}_{\textrm{a}}(\bp,\eps_k){\cal G}_{\textrm{a}}(\bp,\eps_{k+n}){\cal G}_{\textrm{a}}(\bq+\bp,\omega_l-\eps_k){\cal G}_{\textrm{a}}(\bp+\bp,\omega_l-\eps_{k+n}).
\end{split}
\en
Here we have already performed summations over internal band indices and introduced 
\beg\label{LamC}
\lambda(\bq;,\veps_1,\veps_2)=\left(\frac{1}{\tau_{\textrm{t}}}\right)\frac{\theta(-\veps_1\veps_2)}{\left(\varGamma_\pi+|\veps_1-\veps_2|+{\cal D} \bq^2\right)}.
\en
The next step in the process is the evaluation of the corresponding Matsubara sums followed by an analytic continuation. 
As customary, we split MT diagram into a regular and an anomalous contributions \cite{LarkinVarlamovBook},
\beg\label{sigmaMT}
\delta\sigma^{\textrm{MT}}=\delta\sigma_{\textrm{reg}}^{\textrm{MT}}+\delta\sigma_{\textrm{anom}}^{\textrm{MT}}.
\en
These two contributions originate from the different combinations of the poles in the momentum integral in Eq. (\ref{FirstMT}). For the regular contribution we find
\beg\label{sigmaMTregFin}
\delta\sigma_{\textrm{reg}}^{\textrm{MT}}(T)=\left(\frac{e^2\nu_F{\cal D}}{2\pi^2 T}\right)
\int\frac{d^2\bq}{(2\pi)^2}
\int\limits_{-\infty}^{+\infty}\frac{d\omega}{4\pi iT}\coth\frac{\omega}{2T}B''(\bq,\omega)L^R(\bq,\omega). 
\en
In contrast, the anomalous Maki-Thompson correction has different analytical structure. Specifically, we found the following expression:   
\beg\label{sigmaMTanom}
\delta\sigma_{\textrm{anom}}^{\textrm{MT}}(T)=(e^2\nu_F{\cal D})\int \frac{d^2\bq}{(2\pi)^2}\frac{1}{\left(\varGamma_\pi+{\cal D}\bq^2\right)}
\int\limits_{-\infty}^{+\infty}\frac{d\omega}{4\pi T}\frac{[B(\bq,\omega)-B(\bq,-\omega)]}{\sinh^2\frac{\omega}{2T}}L^R(\bq,\omega).  
\en
These expressions can be further analyzed in the limit of low temperatures $T\ll T_{c0}$. We find that the regular contribution survives in the $T\to0$ limit whereas an anomalous contribution vanishes quadratically with $T$. Another nuance is that regular term is weakly (logarithmically) divergent in the ultraviolet of momentum integration. This spurious divergence is due to inapplicability of the diffusion approximation as we are bound by the condition $ql\lesssim1$, where $l$ is the disorder mean-free path. In Fig. \ref{Fig-ResMT} we present the temperature dependence of the regular, anomalous, and total Maki-Thompson contribution.
Table \ref{Table-MT} summarizes main results for the MT term and Appendix \ref{App-MT} provides further technical details of derivations.  

\begin{table}[h!]
\centering
\begin{tabular}{ |c|c|c|} 
 \hline
 $\delta\sigma^{\text{MT}}/T$ & $T\lesssim\Delta_{\text{QCP}}\ll T_{c0}$ & $\Delta_{\text{QCP}}\lesssim T\ll T_{c0}$  \\ 
 \hline\hline
 $\delta\sigma^{\text{MT}}_{\text{reg}}$ & $-\frac{e^2}{8\pi^2}\ln\ln\frac{1}{T_{c0}\tau_\pi}$ & $-2e^2\frac{T}{\varGamma_c}\ln\frac{\varGamma_c}{\Delta_{\text{QCP}}}$  \\ 
 \hline
 $\delta\sigma^{\text{MT}}_{\text{anom}}$ & $\frac{\pi e^2}{3}\frac{T^2}{\varGamma_c\Delta_{\text{QCP}}}$ & $\frac{e^2}{2\pi}\frac{T}{\varGamma_c}\ln\frac{\varGamma_c}{\Delta_{\text{QCP}}}$ \\ 
 \hline
\end{tabular}
  \caption{Asymptotic expressions for the leading temperature dependence of the regular and anomalous MT contributions in a crossover from quantum-to-thermal regimes of fluctuations as extracted from Eqs. \eqref{sigmaMTregFin}--\eqref{sigmaMTanom}.}
 \label{Table-MT}
\end{table}

\begin{figure}[t]
  \centering
 \includegraphics[width=3.5in]{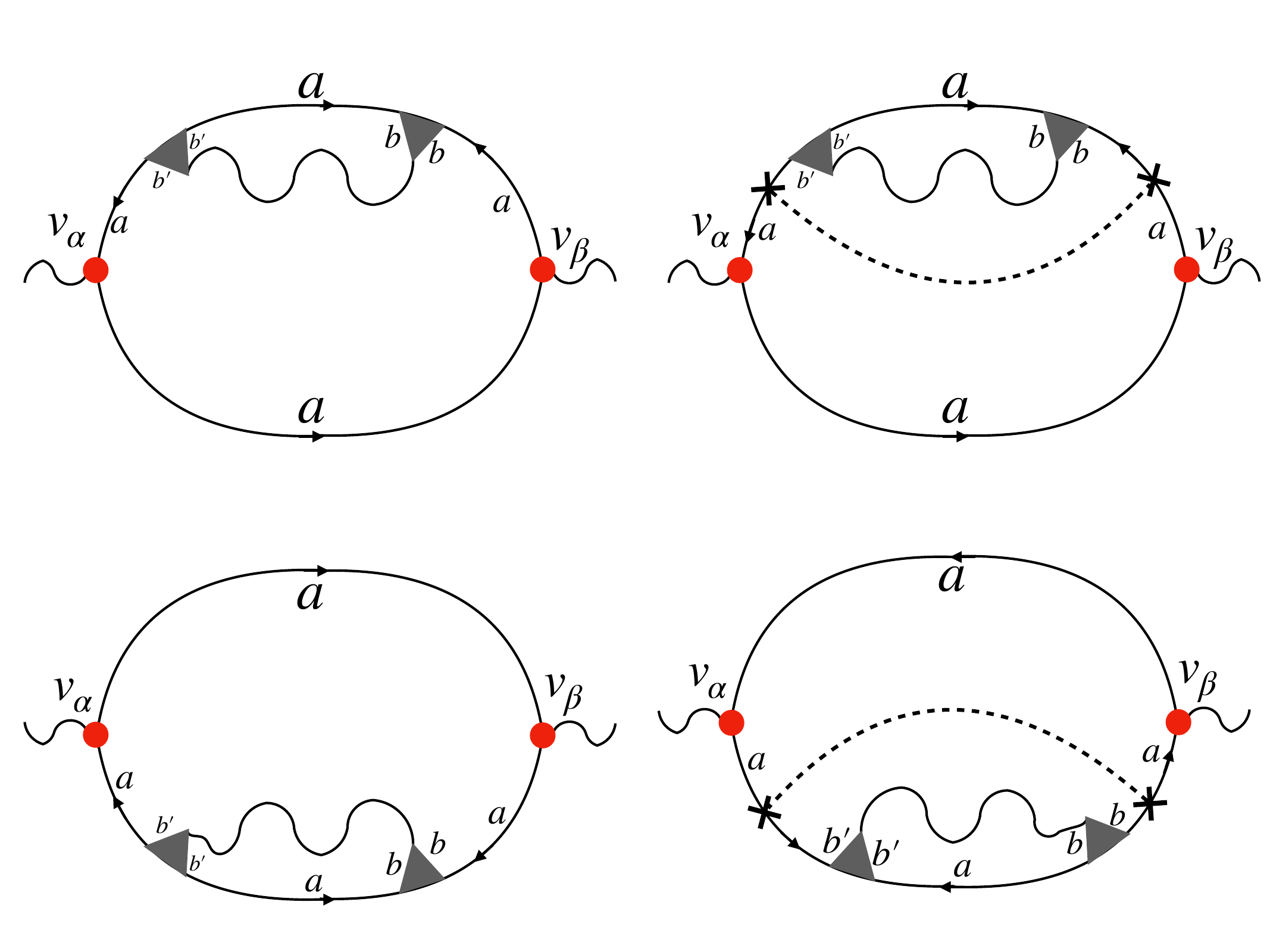}
  \caption{Density-of-states diagrams for the fluctuation correction to conductivity at temperatures above the critical temperature. The Latin letters designate the band indices. Solid lines are single-particle propagators, wavy lines are pairing fluctuation propagators, shaded triangles are Cooperon vertex functions, and solid triangles are the velocity operators. The dashed lines are the single impurity lines, which account for both intra- and interband impurity scattering.}
  \label{Fig4DOS}
\end{figure}
\begin{figure*}[t]
\includegraphics[width=0.325\linewidth]{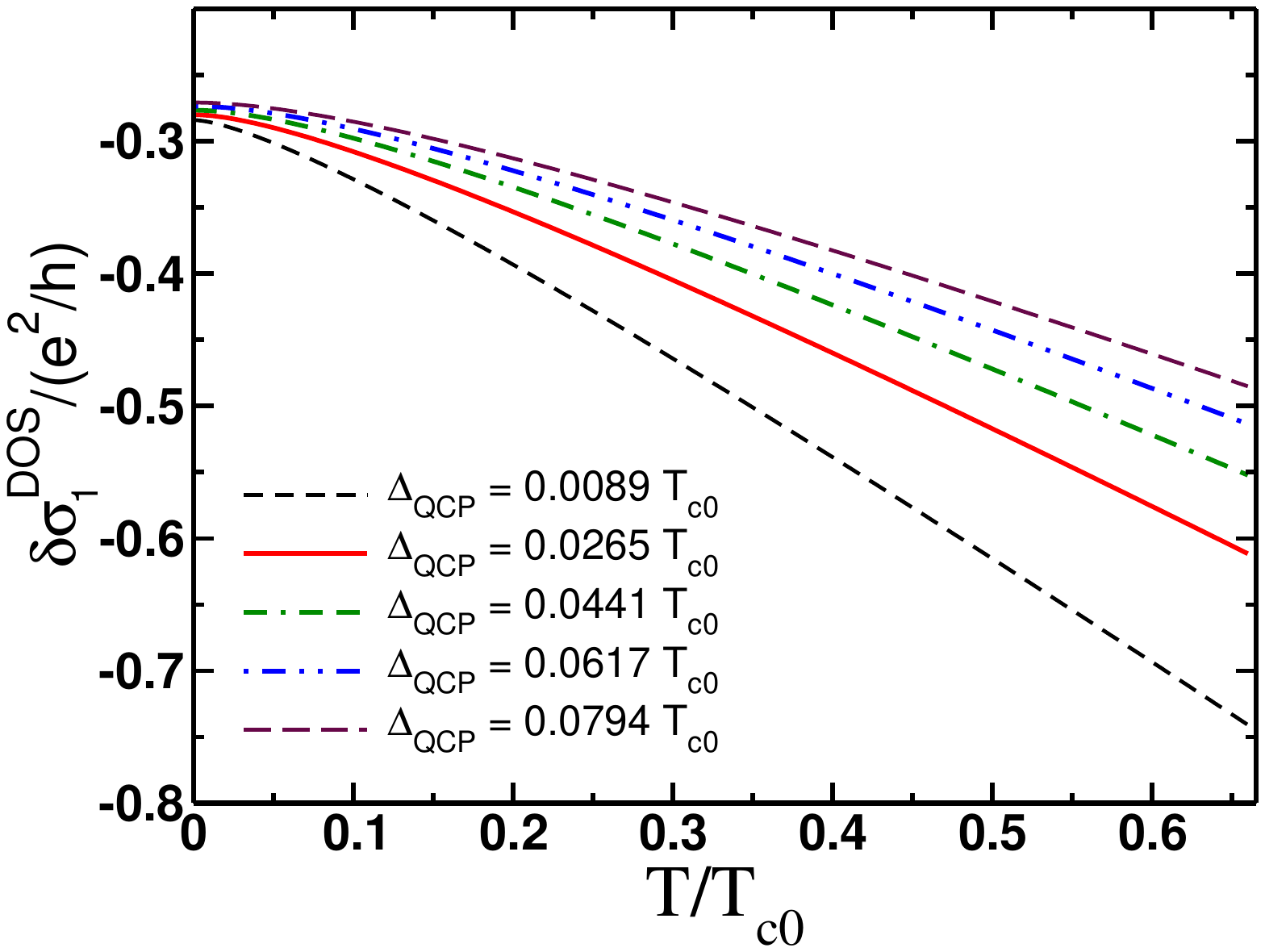}
\includegraphics[width=0.325\linewidth]{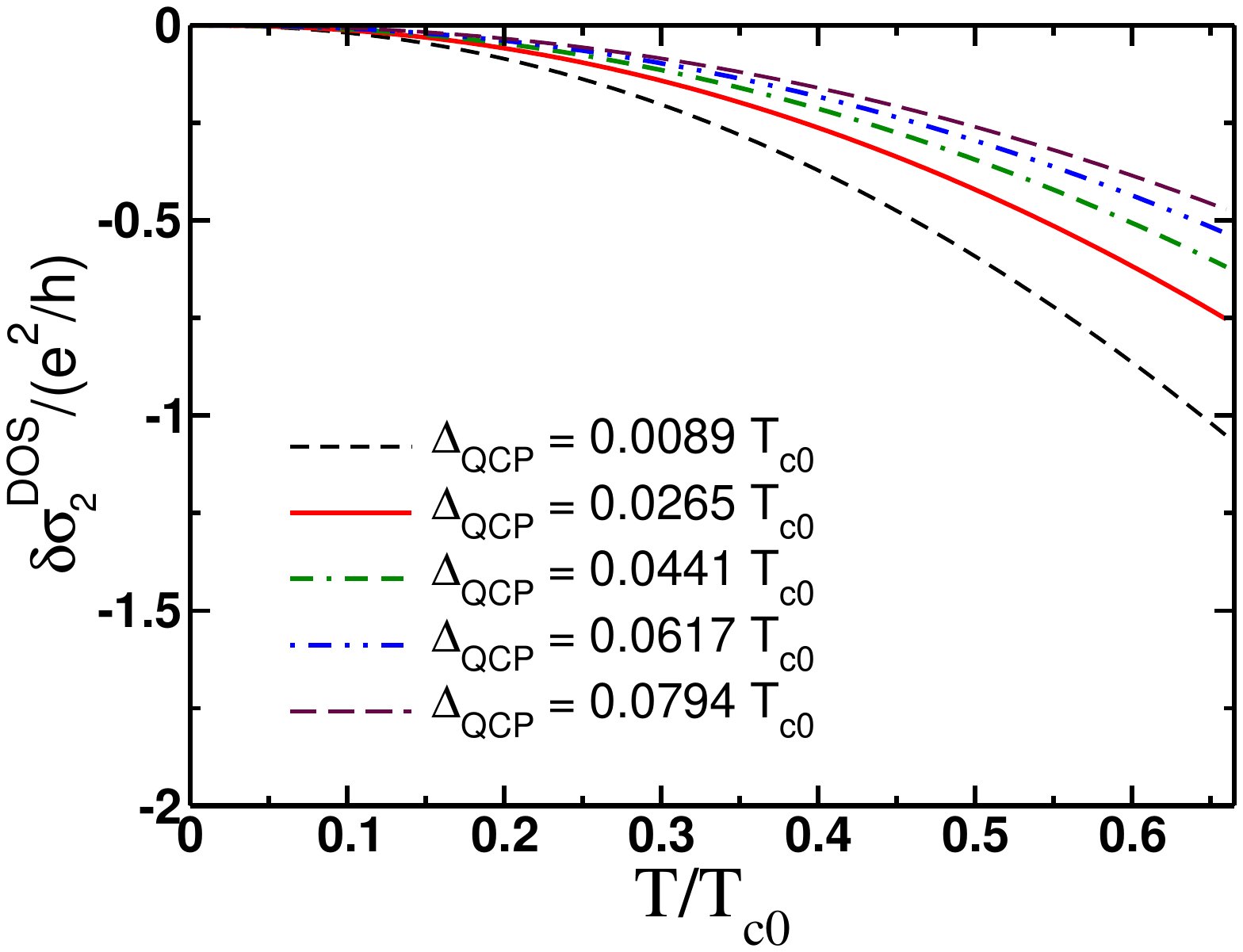}
\includegraphics[width=0.327\linewidth]{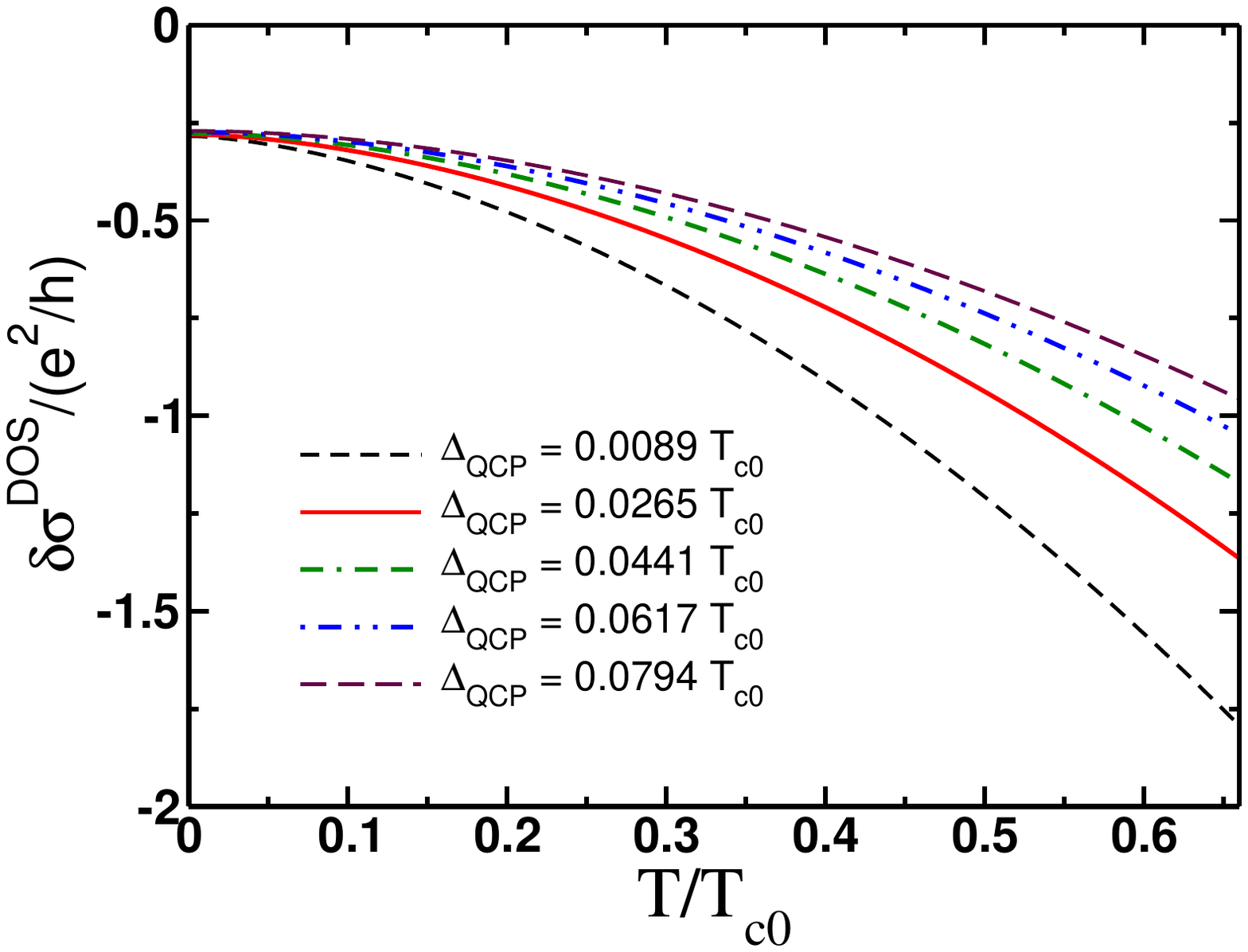}
\caption{Temperature dependence of the fluctuation correction to conductivity from the density-of-states (DOS) diagram, Fig. \ref{Fig4DOS}. The first two panels display the contributions from the first and second terms correspondingly. The third panel shows the total DOS contribution. Comparing these results with the results of the analytic calculations shown in Table \ref{Table-DOS} we note that the crossover temperature $T^*$ separating quantum and thermal regimes is approximately $T^*\approx 0.15 T_{c0}$.}
\label{Fig-ResDOS}
\end{figure*}

\subsection{Density-of-states correction}

We proceed to compute the corrections from the diagrams shown in Fig. \ref{Fig4DOS}. These diagrams account for the fluctuation effects in the single-particle density of states (DOS). As for the two cases above, the internal summation over the band indices produces the expressions for the conductivity correction, which is identical to those found for the single-band superconductor with paramagnetic disorder. The response function, which describes the contributions from the diagrams without impurity line is
\beg\label{KDOS}
K^{\textrm{DOS}}(i\omega_n)=2e^2\sum\limits_{\textrm{a}=c,f}T\sum\limits_{\omega_l}\int\frac{d^2\bq}{(2\pi)^2}\Sigma_{\textrm{a}}^{\textrm{DOS}}(\bq;i\omega_l,i\omega_n)L(\bq,i\omega_l).
\en
Here the factor of two in front of the sum appears due to spin degeneracy. Function $\Sigma_{\textrm{a}}^{\textrm{DOS}}(\bq;i\omega_l,i\omega_n)$ is given by 
\beg\label{SigmaDOS}
\begin{split}
&\Sigma_{\textrm{a}}^{\textrm{DOS}}(\bq;i\omega_l,i\omega_n)=T\sum\limits_{i\eps_k}\lambda^2(\bq;\eps_k,\omega_l-\eps_k)\\ 
&\times\int\frac{d^2\bp}{(2\pi)^2}v_x^2(\bp)\left[{\cal G}_{\textrm{a}}(\bp,\eps_k)\right]^2{\cal G}_{\textrm{a}}(\bp,\eps_k+\omega_n){\cal G}_{\textrm{a}}(\bq-\bp,\omega_l-\eps_k).
\end{split}
\en
Here we have performed summations over internal band indices and function $\lambda$ is defined in Eq. (\ref{LamC}). Similar expression can be also written for the remaining two diagrams with an additional impurity line. 
\begin{table}[h!]
\centering
\begin{tabular}{ |c|c|c|} 
 \hline
 $\delta\sigma^{\text{DOS}}/T$ & $T\lesssim\Delta_{\text{QCP}}\ll T_{c0}$ & $\Delta_{\text{QCP}}\lesssim T\ll T_{c0}$  \\ 
 \hline\hline
 $\delta\sigma^{\text{DOS}}_{\text{1}}$ & $-\frac{e^2}{8\pi^2}\ln\ln\frac{1}{T_{c0}\tau_\pi}$ & $-2e^2\frac{T}{\varGamma_c}\ln\frac{\varGamma_c}{\Delta_{\text{QCP}}}$  \\ 
 \hline
 $\delta\sigma^{\text{DOS}}_{\text{2}}$ & ${-\frac{2e^2}{3}\frac{T^2}{\varGamma_c\Delta_{\textrm{QCP}}}}$ & {-${e^2}\frac{T}{\varGamma_c}\ln\frac{\varGamma_c}{\Delta_{\text{QCP}}}$}  \\ 
 \hline
\end{tabular}
  \caption{Asymptotic expressions for the leading temperature dependence of the first and second DOS contributions in a crossover from quantum-to-thermal regimes of fluctuations as extracted from Eqs. \eqref{sigma1DOS}--\eqref{sigma2DOS}.}
 \label{Table-DOS}
\end{table}

The remaining steps in the calculation of the correction to conductivity are essentially identical to the ones used to compute the Maki-Thompson correction. 
Specifically, here we also find two distinct contributions to conductivity, 
\beg\label{sigmaDOS}
\delta\sigma^{\textrm{DOS}}=\delta\sigma_{1}^{\textrm{DOS}}+\delta\sigma_{2}^{\textrm{DOS}}.
\en
For the first one we find the an expression that coincides with that of a regular MT formula given in Eq. (\ref{sigmaMTregFin}), 
\beg\label{sigma1DOS}
\delta\sigma_{1}^{\textrm{DOS}}(T)=\delta\sigma^{\text{MT}}_{\text{reg}}
\en
The expression for the remaining correction to conductivity is different and reads
\beg\label{sigma2DOS}
\delta\sigma_{2}^{\textrm{DOS}}(T)=\frac{e^2{\cal D}}{4\pi T}\int\frac{d^2\bq}{(2\pi)^2}\int\limits_{-\infty}^{+\infty}\frac{d\omega}{2\pi T}
\frac{[B'(\bq,\omega)-B'(\bq,-\omega)]}{\sinh^2\left(\frac{\omega}{2T}\right){\left[\ln({T}/{T_{c0}})+\varPsi(\bq,-i\omega)\right]}}.
\en
The details of our analysis of these two contributions in quantum $(T\ll \Delta_{\textrm{QCP}})$ and thermal $(T\gg \Delta_{\textrm{QCP}})$ regimes are provided in Appendix C. 
In Fig. \ref{Fig-ResDOS} we present the temperature dependence for the individual $\delta\sigma^{\text{DOS}}_{1,2}$ and total DOS contribution to conductivity.
The summary of the results is presented in Table \ref{Table-DOS}.

\subsection{Diffusion constant renormalization}
\begin{figure}[t]
  \centering
 \includegraphics[width=3.55in]{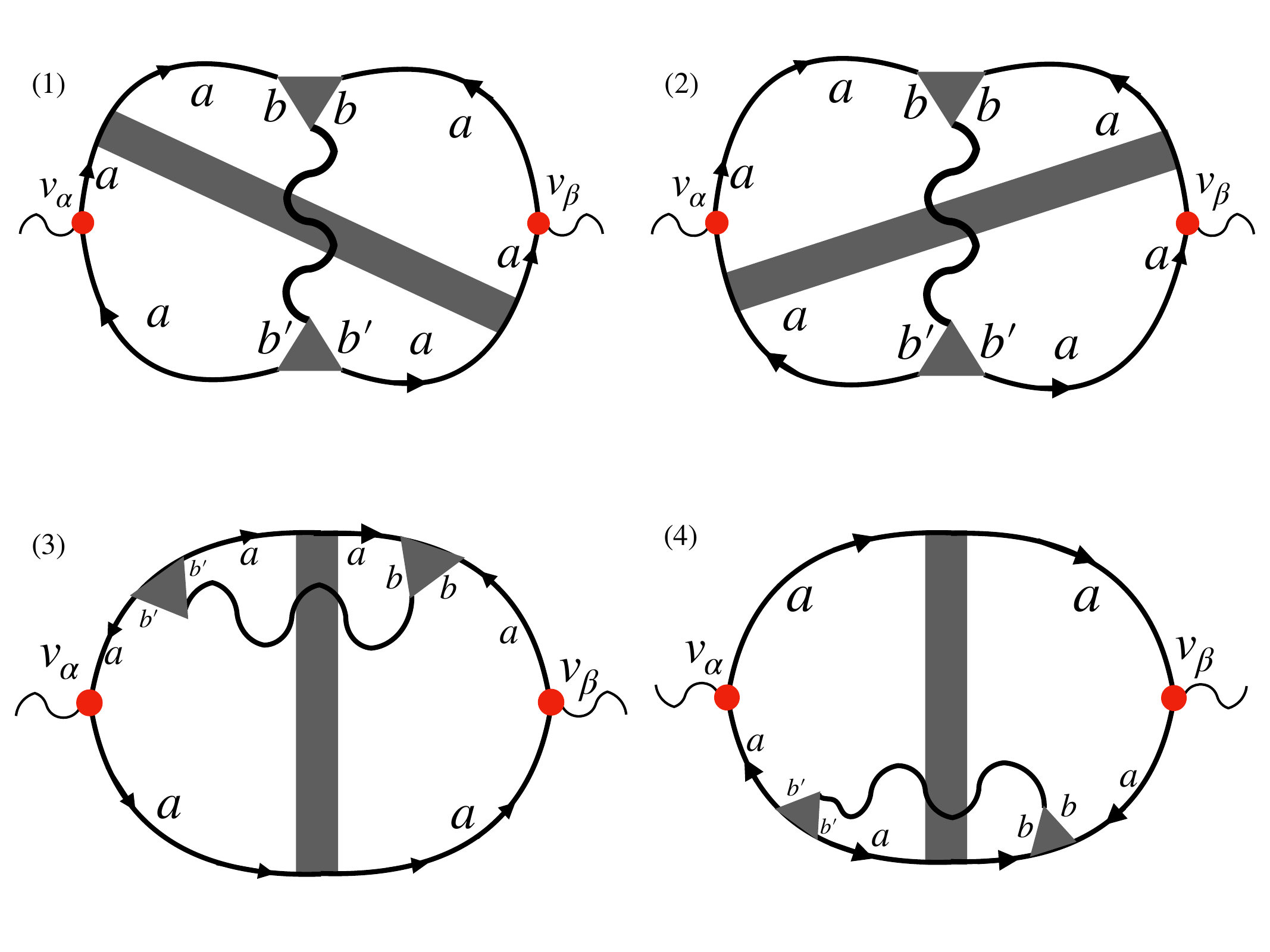}
  \caption{Feynman diagrams representing the fluctuation corrections to conductivity. Due to their distinct contributions in both quantum and thermal fluctuation regimes, these diagrams must be considered separately from the Maki-Thompson and density-of-states diagrams. They describe the contributions to conductivity due to renormalization of the diffusion coefficient (DCR). The Roman letters denote the band indices, solid black lines stand for the single-particle propagators, wavy line denotes the fluctuation propagator, shaded triangle designates the diagonal part of the Cooperon function ${\cal C}_{\textrm{aa}}$, and shaded triangles denote the impurity ladders including the intra- and interband disorder scattering (see text).}
  \label{FigDCR}
\end{figure}

In this section we discuss the corrections to dc conductivity given by the four diagrams shown in Fig. \ref{FigDCR}. In all expressions below if we do not explicitly comment on the origin of the specific quantities, that means they have already been defined in the main text. 

Since the first two diagrams give identical contributions, we begin our discussion with the first diagram. For this diagram the current-current correlation function has the following analytical expression:
\beg\label{DCR1}
\begin{split}
&K_{1}^{\textrm{DCR}}(i\omega_\nu)=4e^2T^2\sum\limits_{\omega_k}\sum\limits_{\veps_n}\int\frac{d^2\bq}{(2\pi)^2}\lambda(\bq;\veps_n,\omega_k-\veps_n)\lambda(\bq;\veps_n+\omega_\nu,\omega_k-\veps_n-\omega_\nu)\\&\times {\cal C}(\bq;\veps_n+\omega_\nu,\omega_k-\veps_n)L(\bq,\omega_k)\int\frac{d^2\bp}{(2\pi)^2}v_x(\bp){\cal G}(\bp,\veps_n){\cal G}(\bp,\veps_n+\omega_\nu){\cal G}(\bq-\bp,\omega_k-\veps_n)\\&\times\int\frac{d^2\bp'}{(2\pi)^2}v_x(\bq-\bp'){\cal G}(\bp',\veps_n+\omega_\nu){\cal G}(\bq-\bp',\omega_k-\veps_n-\omega_\nu){\cal G}(\bq-\bp',\omega_k-\veps_n).
\end{split}
\en
In this expression we have already performed the summations over the band indices. The new quantity, which appears here, ${\cal C}(\bq;\veps_n+\omega_\nu,\omega_k-\veps_n)$, represents the Cooperon block (shaded rectangle in Fig. \ref{FigDCR}), 
\beg\label{Coop}
{\cal C}(\bq;\veps_1,\veps_2)=\frac{1}{2\pi\nu_F\tau_{\textrm{t}}^2}\left(\frac{\theta(-\veps_1\veps_2)}{\varGamma_\pi +|\veps_1-\veps_2|
+{\cal D}\bq^2}+\frac{\theta(-\veps_1\veps_2)}{|\veps_1-\veps_2|+{\cal D}\bq^2}\right).
\en
We immediately note that the expression for the Cooperon has two contributions: one which contains the pair-breaking rate, and another one which does not. 

The calculation proceeds with the computation of the momentum integrals followed up by the summation over the fermionic Matsubara frequencies $\veps_n$ and subsequent analytic continuation to real frequencies $\omega_\nu \to -i\omega$. As a result, we find the following expression for the conductivity correction, which we will split into the sum of distinct contributions: 
\beg\label{sigma1DCR}
\delta\sigma^{\textrm{DCR}}=\delta\sigma_1^{\textrm{DCR}}+\delta\sigma_2^{\textrm{DCR}}+\delta\sigma_3^{\textrm{DCR}}+\delta\sigma_4^{\textrm{DCR}}.
\en
The first contribution originates from the first term in the expression for the Cooperon (\ref{Coop}), which contains the pair-breaking rate $\varGamma_\pi$ and is given by 
\beg\label{dsigmaDCRpi}
\delta\sigma_{1,\pi}^{\textrm{DCR}}=\frac{\sigma_{D}}{(4\pi)^3 T^2}\sum\limits_{k=-\infty}^\infty\int\frac{{\cal D}q_x^2d^2\bq}{(2\pi)^2}L(\bq;|\omega_k|)\psi'''\left(\frac{1}{2}+\frac{\varGamma_\bq+|\omega_k|}{4\pi T}\right).
\en
Here $\sigma_D=2e^2\nu_F{\cal D}$ is Drude conductivity and $\varGamma_\bq$ was introduced earlier inn Eq. \eqref{Defxq}. Expression for the second contribution to conductivity reads
\beg\label{dsigmaDCR0}
\begin{split}
&\delta\sigma_{1,0}^{\textrm{DCR}}=\frac{\sigma_{D}}{(4\pi)^3 \varGamma_\pi^2}\sum\limits_{k=-\infty}^\infty\int\frac{{\cal D}q_x^2d^2\bq}{(2\pi)^2}L(\bq;|\omega_k|)\\
&\times\left\{\psi'\left(\frac{1}{2}+\frac{{\cal D}\bq^2+|\omega_k|}{4\pi T}\right)-\psi'\left(\frac{1}{2}+\frac{\varGamma_\bq+|\omega_k|}{4\pi T}\right)+\left(\frac{\varGamma_\pi}{4\pi T}\right)\psi''\left(\frac{1}{2}+\frac{\varGamma_\bq+|\omega_k|}{4\pi T}\right)\right\}.
\end{split}
\en
The remaining step is to perform the analytic continuation with respect to bosonic frequencies $\omega_k\to -i\omega$, which, in the case of the expressions above, is straightforward. In Appendix D we will analyze each of these two contributions in two separate regimes: (i) quantum fluctuations regime, when $T\ll \Delta_{\textrm{QCP}}$ ($\Delta_{\textrm{QCP}}\equiv|\varGamma_\pi-\varGamma_\pi^{(c)}|$) and (ii) thermal fluctuations regime when $T\gg\Delta_{\textrm{QCP}}$.

Finally, let us compute the contribution from the remaining two diagrams in Fig. \ref{FigDCR}.
Analytical expression for the third diagram is
\beg\label{DCR1}
\begin{split}
&K_{3}^{\textrm{DCR}}(i\omega_\nu)=4e^2T^2\sum\limits_{\omega_k}\sum\limits_{\veps_n}\int\frac{d^2\bq}{(2\pi)^2}\lambda(\bq;\veps_{n+\nu},\omega_k-\veps_{n+\nu})\lambda(\bq;\veps_{n+\nu},\omega_k-\veps_{n+\nu})\\&\times {\cal C}(\bq;\veps_n,\omega_k-\veps_{n+\nu})L(\bq,\omega_k)\int\frac{d^2\bp}{(2\pi)^2}v_x(\bp){\cal G}(\bp,\veps_{n+\nu}){\cal G}(\bp,\veps_n)\\&\times\int\frac{d^2\bp'}{(2\pi)^2}v_x(\bp'){\cal G}(\bp',\veps_{n+\nu}){\cal G}(\bq-\bp',\omega_k-\veps_{n+\nu}){\cal G}(\bp',\veps_n).
\end{split}
\en
Here we adopted the shorthand notation $\veps_{n+\nu}=\veps_n+\omega_\nu$ for brevity.

\begin{table}[h!]
\centering
\begin{tabular}{ |c|c|c|} 
 \hline
 $\delta\sigma^{\text{DCR}}/T$ & $T\lesssim\Delta_{\text{QCP}}\ll T_{c0}$ & $\Delta_{\text{QCP}}\lesssim T\ll T_{c0}$  \\ 
 \hline\hline
 $\delta\sigma_{1+2}^{\text{DCR}}$ & $ \frac{e^2}{8\pi^2}\ln\ln\frac{1}{T_{c0}\tau_\pi}$ & {$\frac{e^2}{8\pi^2}\frac{T}{\varGamma_c}$}  \\ 
 \hline
 $\delta\sigma^{\text{DCR}}_{3+4}$ & ${\frac{e^2}{24\pi^2}\ln\ln\frac{1}{T_{c0}\tau_\pi}}$ & {$\frac{e^2}{24\pi^2}\frac{T}{\varGamma_c}$}  \\ 
 \hline
\end{tabular}
  \caption{Asymptotic expressions for the leading temperature dependence of the first and second DOS contributions in a crossover from quantum-to-thermal regimes of fluctuations as extracted from Eqs. \eqref{sigma1DOS} and \eqref{sigma2DOS}.}
 \label{Table-DCR}
\end{table}

\begin{figure*}[t]
\includegraphics[width=0.75\linewidth]{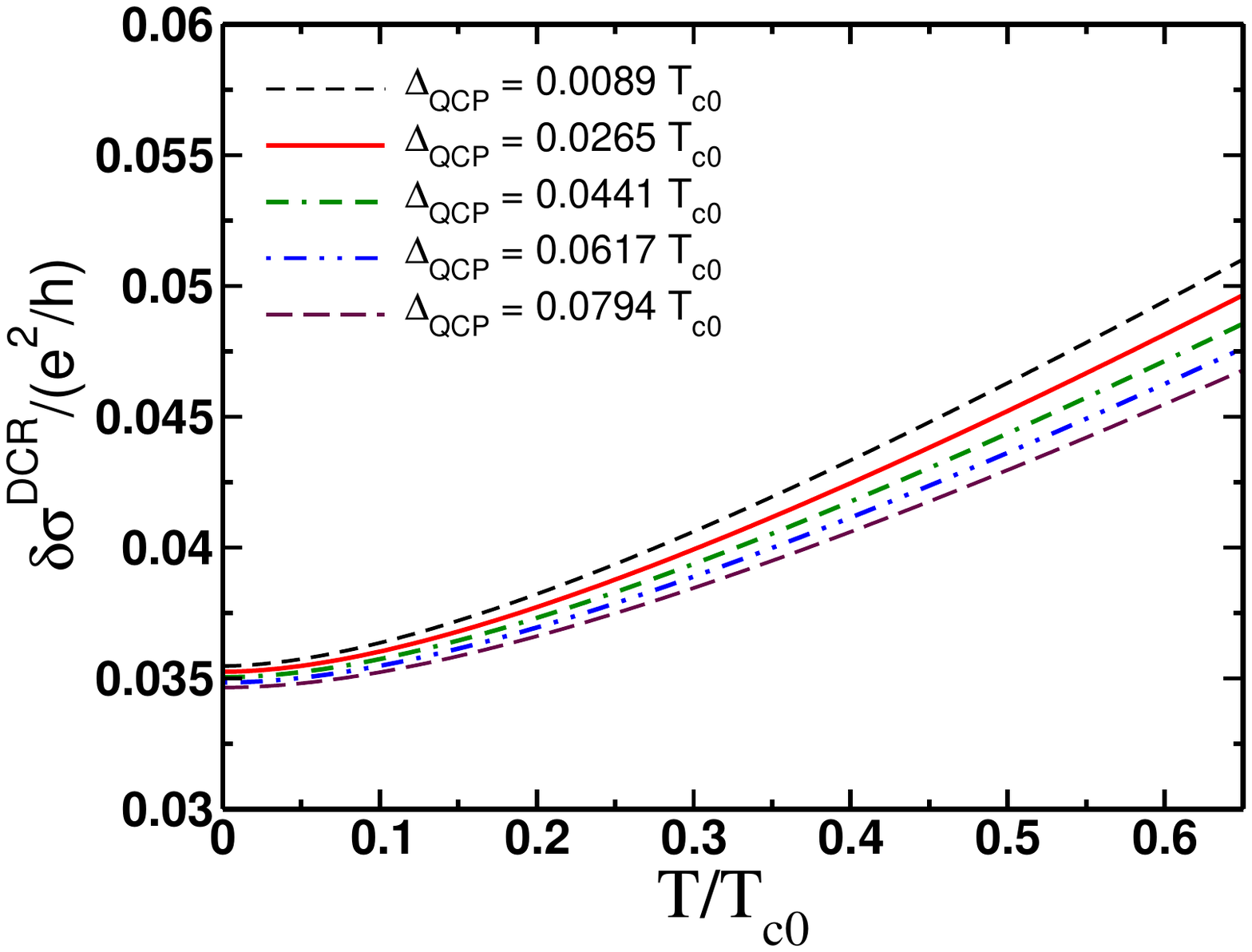}
\caption{Temperature dependence of the total quantum correction to conductivity described by the DCR diagrams, Fig. \ref{FigDCR}.}
\label{Fig-DCR}
\end{figure*}

We have determined that the second term in the Cooperon, Eq. (\ref{Coop}), gives negligible contribution to the conductivity correction from the first two diagrams, i.e., $\delta\sigma_{1,\pi}^{\textrm{DCR}}\gg\delta\sigma_{1,0}^{\textrm{DCR}}$. It turns out the same relation holds for the contributions from the remaining two diagrams and in what follows we provide the formulas found for the first part of the Cooperon only. Subsequent integration over momenta followed up by the summation over the fermionic Matsubara frequencies yields the following expression, which we represent as a sum of two terms,
\beg\label{Qxx34}
\begin{split}
&K_{3}^{\textrm{DCR}}=\sigma_D\left(\frac{T}{\omega_\nu^2}\right)\sum\limits_{\omega_k}\int\frac{{\cal D}q_x^2d^2\bq}{(2\pi)^2}L(\bq,|\omega_k|)
\left[\psi\left(\frac{1}{2}+\frac{\varGamma_\bq+|\omega_k|}{4\pi T}\right)\right.\\&\left.-\psi\left(\frac{1}{2}+\frac{\varGamma_\bq+\omega_\nu+|\omega_k|}{4\pi T}\right)+\frac{\omega_\nu}{4\pi T}\psi'\left(\frac{1}{2}+\frac{\varGamma_\bq+\omega_\nu+|\omega_k|}{4\pi T}\right)\right]\\&+
\sigma_D\left(\frac{T}{\omega_\nu^2}\right)\sum\limits_{\omega_k}\int\frac{{\cal D}q_x^2d^2\bq}{(2\pi)^2}L(\bq,|\omega_k|)\left[\psi\left(\frac{1}{2}+\frac{\varGamma_\bq+\omega_\nu+|\omega_{k+\nu}|}{4\pi T}\right)\right.\\&\left.-\psi\left(\frac{1}{2}+\frac{\varGamma_\bq+|\omega_{k+\nu}|}{4\pi T}\right)-\frac{\omega_\nu}{4\pi T}\psi'\left(\frac{1}{2}+\frac{\varGamma_\bq+|\omega_{k+\nu}|}{4\pi T}\right)\right].
\end{split}
\en
The procedure of the analytic continuation for the first of these two terms is straightforward. As for the second one, although it is more involved, it is technically similar to the one performed for the Aslamazov-Larkin correction. Following the analytic continuation procedure with the expansion of the kernel $K_{3}^{\textrm{DCR}}$ in powers of $\omega_\nu\to -i\omega$ we find that the conductivity correction from that second term vanishes identically and we are left with the correction originating from the first term in (\ref{Qxx34}). Specifically, we find
\beg\label{dsigmaDCRDOS}
\delta\sigma_{3,\pi}^{\textrm{DCR}}(T)=\frac{\sigma_{D}}{3(4\pi)^3 T^2}\sum\limits_{k=-\infty}^\infty\int\frac{{\cal D}q_x^2d^2\bq}{(2\pi)^2}L(\bq;|\omega_k|)\psi'''\left(\frac{1}{2}+\frac{\varGamma_\bq+|\omega_k|}{4\pi T}\right).
\en
We immediately notice that this correction is a factor of three smaller than (\ref{dsigmaDCRpi}). We compute the temperature dependence of all four contributions numerically, Fig. \ref{Fig-DCR}. An asymptotic analysis of the expressions above can be performed similarly to how it has been done for other contributions. The details of this analysis are given in Appendix D and the results are shown in Table \ref{Table-DCR}. 

Note that the DCR corrections are positive both in thermal and quantum regimes and are weakly ultraviolet divergent in the quantum regime.  Now, if we compare all the results in Tables \ref{Table-AL}, \ref{Table-MT}, \ref{Table-DOS}, and \ref{Table-DCR}, we conclude that in the quantum regime both MT and DOS contributions are dominant rendering the quantum correction to conductivity to be negative, while in the thermal regime the Aslamazov-Larkin contribution is a leading one and so the corresponding correction is positive. 

\section{Summary and discussion}\label{sec:Discussion}

In this paper, we presented a comprehensive microscopic analysis of the leading fluctuation-induced corrections to the conductivity of a disordered multiband metal in the vicinity of the superconducting quantum critical point. In contrast to previous considerations of this problem in different models belonging to the same universality class, which were based on the effective bosonic field theories, we carried out our analysis based on the direct diagrammatic approach, which enables us to retain both fermionic quasiparticles and bosonic modes on equal footing. We find that in the quantum limit of fluctuations the theory is regular in the infrared, in other words as the system approaches the superconducting state from the metallic side, its transport properties do not reflect the proximity of another phase. The corrections to conductivity are regular and practically universal $\sim e^2/h$. We find only extremely weak double-logarithmic divergence in the ultraviolet, which is a simple reflection of the inapplicability of the diffusive limit when the theory is extrapolated to large momenta. The natural cut-off of fluctuating bosonic modes is set by $\bq_{\text{max}}l\sim1$. As a result, the total conductivity correction is negative in $T\to0$ limit, namely  
\begin{equation}
\delta\sigma_{\text{tot}}(T=0)=-\frac{e^2}{12\pi^2}\ln\ln\frac{1}{T_{c0}\tau_\pi}. 
\end{equation}
This is a peculiar result as this localizing correction results from the attractive interactions in the Cooper channel. Note that in the classical regime of fluctuations such corrections are always positive \cite{LarkinVarlamovBook}.  
The negative sign is a result of a delicate cancellation between positive corrections to the diffusion constant renormalization, and negative corrections stemming from the density of states effect and regular part of the destructive Maki-Thompson interference.       
At finite temperatures, the overall temperature dependence of $\delta\sigma_{\text{tot}}(T)$ displays the nonmonotonic behavior as depicted in Fig. \ref{Fig-Main1}. It depends on the trajectory toward the QCP. At temperatures exceeding the gap to QCP, the correction is positive and governed by the classical part of the Aslamazov-Larkin term. Curiously, this correction is linear in temperature, see Table \ref{Table-AL} for the summary. 

All the complications and subtleties of the quantum regime of superconducting fluctuations in the context of transport properties can be traced to their dynamical nature as captured by the vertex corrections. This feature can be best described in the example of the Aslamazov-Larkin process. Indeed, the key element of the AL diagram is the triangular block that consists of three Green's functions and two impurity ladders. This block depends on three frequencies: the injected frequency $\Omega$ at which conductivity is calculated, which is set to zero at the end of the calculation to extract the dc limit; the fermionic frequency $\varepsilon$ in the loop of Green's functions; and the bosonic frequency $\omega$ that defines the pair propagator. In the classical regime of fluctuations, $T\ll T-T_c$, the pole structure of the pair propagator set the characteristic frequency of soft bosonic modes to $\{\mathcal{D}\bq^2,\omega\}\sim T-T_c$. At the same time, fermionic fluctuations occur at characteristic energies $\varepsilon\sim T$. Therefore $\varepsilon\gg\omega$ and in the calculation of the triangular block all bosonic frequencies can be dropped thus making it to be purely static. The whole block scales linearly with momentum $\bq$ since it originates from the current vertex. In contrast, near the QCP when $T_c\to 0$, all energies are of the same order, $\omega\sim\varepsilon\sim T$, and this vertex function has a complicated dynamical structure. Upon an analytical continuation, it generates three different vertices in the causal sector (retarded and advanced components) depending on whether energies are on the same or different half-planes with respect to the given branch cut. In the end, expansion over $\Omega$ generates additional contributions to the conductivity that do not have a classical analog. The same argumentation applies to all the contributions.  

We close with a broader perspective on several outstanding problems where an extension of the theory presented in this paper may find useful applications. (i) The applicability of our results is restricted by the perturbation theory in fluctuations. It is controlled by the Ginzburg-Levanyuk parameter, $Gi=1/(\nu_F\mathcal{D})\ll1$, which in the 2D case is determined by the inverse of the dimensionless conductance \cite{LarkinVarlamovBook}. Going beyond the leading one-loop order requires consideration of additional diagrams that describe the interaction of fluctuations. Thus far this analysis was performed only in the classical region of fluctuations \cite{LO-JETP2001,AL-PRB2010} (see also Refs. \cite{VarlamovDorin,Reizer-MT,Reizer-Tunnel,Lerner,AL-PRB2010-Tunnel} where some particular higher-order diagrams were investigated in the context of granular superconducting systems and tunnel junctions). This analysis suggests a crossover scale $T-T_c\sim T_c\sqrt{Gi}$ where nonlinear effects of fluctuations start to dominate. The proper resummation of the most singular contributions remains an outstanding problem (one should notice though that as $T_c$ is suppressed to zero this region narrows). (ii) Throughout the paper we focused on the 2D systems; however, the derived results for the fluctuation corrections to the conductivity apply in 3D as well with the proper replacement of the momentum integrals. We expect weaker singularities in the 3D geometry simply based on the phase space argumentation for fluctuations at long wavelengths. (iii) The next point concerns transport anomalies observed in homogeneously disordered superconducting films and the associated physics of the superconductor-to-insulator transition (SIT). As is known near $T_c$ there is a parametrically small temperature window close to the Berezinskii-Kosterlitz-Thouless temperature $T_{\text{BKT}}$ within which the normal state resistance rapidly drops to exponentially small values. Whereas above $T_c$ the resistance is governed by the superconducting fluctuations, below $T_c$ the resistance is determined by the proliferation of vortex excitations in the order parameter field those dynamics require careful account of a phase relaxation mechanism. The crossover between these regimes is commonly described by the Halperin and Nelson interpolation formula \cite{HalperinNelson1979}. To some extent, this formula can be corroborated in the microscopic field theory \cite{KonigPRB15,KonigPRB21}. The nature of this crossover in the quantum regime when $T_c\to0$ is not understood. It may shed light onto the origin of anomalous metal phases \cite{KapitulnikRMP19} of SIT where film resistance saturates to a constant value. At a bare minimum, our perturbative calculation is suggestive of this possibility. (iv) The final point concerns strongly correlated metal where the interplay between superconducting, spin fluctuation, and Fermi Liquid phases near the quantum critical endpoint leads to a peculiar temperature and field dependence of the resistance, see e.g., Ref. \cite{Paglione}. A proper account of the dynamic nature of these fluctuations in different competing pairing channels may lead to fruitful results and an explanation of observed anomalies in the resistivity of a system. 

\section*{Acknowledgments}

We thank A. Andreev, A. Chubukov, P. Coleman, S. Kivelson, D. Maslov, A. Nevidomskyy, R. Ramazashvili, B. Spivak, A. Varlamov, and C. Varma for useful discussions on various topics related to this study. This work was financially supported by the National Science Foundation Grant No. DMR-2002795 (M.D.) and by the NSF Grant No. DMR-2203411 (A.L.). M. K. acknowledges financial support from the Israel Science Foundation, Grant No. 2665/20. A.L. acknowledges hospitality of the Max Planck Institute for Solid State Research, where this work was performed in part, and a Research Fellowship funded by the Alexander von Humboldt Foundation. Parts of this paper were written during the Aspen Center of Physics 2023 Summer Programs on ``Quantum Materials: Experimental Enigmas and Theoretical Challenges" (A.L. and M.D.) and ``New Directions on Strange Metals in Correlated Systems" (M.D.), which was supported by the National Science Foundation Grant No. PHY-2210452.

\appendix 
\section{Asymptotic expressions for $\delta\sigma^{\text{AL}}$}\label{App-AL}
In this and the following sections, we present the details of the calculation of the asymptotic values for the quantum correction to conductivity in quantum and thermal regimes. In our analysis of the corresponding integrals, we will use several approximations, which are naturally applicable for all four contributions (AL, MT, DOS, and DCR) and so discussion in each section may seem a bit repetitive. Nevertheless, for the sake of the reader's convenience, we will outline key steps in the calculation in each of the four cases separately. 

\paragraph{Quantum critical regime.} In the limit of low enough temperatures $T\tau_{\textrm{t}}\ll 1$ the expressions above can be simplified using the following approximate
expressions for the digamma functions:
\beg\label{GenExpand}
\begin{split}
&\psi\left(\frac{1}{2}+\frac{x_\bq-i\xi}{2}\right)\approx\ln\left(\frac{x_\bq-i\xi}{2}\right), \quad
\psi'\left(\frac{1}{2}+\frac{x_\bq-i\xi}{2}\right)\approx\frac{2}{x_\bq-i\xi}.
\end{split}
\en
Then for the first two corrections to Aslamazov-Larkin contribution from the expansion of the fluctuation propagator one obtains
\beg\label{sigma1pp2}
\begin{split}
&\delta\sigma_1^{\textrm{AL}}\approx\frac{e^2}{4\pi}\int\limits_0^\infty\frac{y_\bq dy_\bq}{(2\pi T)^2}\int\limits_{-\infty}^\infty \frac{d\xi}{\sinh^2(\pi \xi)}\frac{x_\bq^4\xi^2}{(x_\bq^2+\xi^2)^2[x_\bq^2\ln^2(x_\bq/x_c)+\xi^2]^2}, \\
&\delta\sigma_2^{\textrm{AL}}=-\frac{e^2}{\pi^2}\int\limits_0^\infty\frac{y_\bq dy_\bq}{(2\pi T)^2}\int\limits_{-\infty}^\infty\frac{\coth(\pi\xi)\xi d\xi}{(x_\bq^2+\xi^2)^3}\textrm{Re}\left\{\frac{(i\xi+x_\bq)(i\xi+2x_\bq)}{\ln^3\left(\frac{x_\bq-i\xi}{x_c}\right)}\right\},
\end{split}
\en
Here we are using the following dimensionless notations: $x_\bq=x_c+y_\bq/2\pi T$, $x_c=\varGamma_c/2\pi T$, $\xi=\omega/2\pi T$, and $y_\bq={\cal D}\bq^2$.  

The integral over $\xi$ in the expression for $\delta\sigma_1^{\textrm{AL}}$ can be computed approximately employing the diffusive approximation $T\tau_{\textrm{t}}\ll 1$, which yields 
\beg\label{sigma1ALint}
\delta\sigma_1^{\textrm{AL}}\approx\frac{e^2}{6\pi}(2\pi T)^2\int\limits_0^\infty\frac{y_\bq dy_\bq}{(\varGamma+y_\bq)^4\ln^4(x_\bq/x_c)}.
\en
The main contribution to the remaining integral over $y_\bq$ comes from the region of small values of $y_\bq$ and so we readily obtain the following estimate:
\beg\label{dAL1QR}
\delta\sigma_1^{\textrm{AL}}(T\ll\Delta_{\textrm{QCP}})\approx\frac{e^2}{6\pi}\frac{T^2}{\Delta_{\textrm{QCP}}^2}.
\en

The analysis  of the temperature dependence of $\delta\sigma_2^{\textrm{AL}}$ (including a numerical factor) can be somewhat simplified if one makes an additional change of variable $\xi=x_c\eta$. Then for the first two corrections to Aslamazov-Larkin contribution from the expansion of the fluctuation propagator it obtains
\beg\label{sigmap2qr}
\begin{split}
&\delta\sigma_2^{\textrm{AL}}=-\frac{e^2}{\pi^2}\int\limits_0^\infty\frac{y_\bq dy_\bq}{\varGamma_c^2}\int\limits_{-\infty}^\infty\frac{\coth(\pi x_c\eta)\eta d\eta}{(x_\bq^2/x_c^2+\eta^2)^3}\textrm{Re}\left\{\frac{(i\eta+x_\bq/x_c)(i\eta+2x_\bq/x_c)}{\ln^3\left({x_\bq/x_c-i\eta}\right)}\right\}.
\end{split}
\en
Note that the only temperature-dependent term in this expression is an argument of the $\coth$. 
Clearly, in the quantum regime $T\ll|\Delta_{\textrm{QCP}}|$, apart from the region when $\eta\ll 1$ the argument of the $\coth$ is much larger than one and, therefore, we can approximate $\coth(\pi x_c\eta)\approx \textrm{sign}\eta$. Consequently, as a result, the temperature dependence completely drops out from this expression. The remaining integral can be easily computed and we get
\beg\label{sigmap2qrapp}
\delta\sigma_2^{\textrm{AL}}(T\ll\Delta_{\textrm{QCP}})\approx\left(\frac{33\pi^2-316}{32\pi^2}\right)e^2.
\en

Now we consider the remaining two contributions to the AL correction, which can be written in the following form:
\beg\label{sigma2p2ppfin}
\begin{split}
&\delta{\sigma}_3^{\textrm{AL}}=-\frac{2e^2}{\pi}\left(\frac{{\cal D}}{\pi T}\right)^2\left(\frac{1}{\pi T\tau_\pi^{(c)}}\right)^2\int\frac{q_x^2d^2\bq}{(2\pi)^2}
\textrm{Im}\left\{\int\limits_{-\infty}^\infty\frac{\coth(\pi\xi)d\xi}{(x_\bq-i\xi)^3(\tilde{x}_\bq-i\xi)^2}\right\}, \\
&\delta\sigma_4^{\textrm{AL}}=-\frac{2e^2}{\pi}\left(\frac{{\cal D}}{\pi T}\right)^2\left(\frac{1}{\pi T\tau_\pi^{(c)}}\right)^2\int\frac{q_x^2d^2\bq}{(2\pi)^2}
\int\limits_{-\infty}^\infty\frac{x_\bq^2\coth(\pi\xi)\xi d\xi}{(x_\bq^2+\xi^2)^3(\tilde{x}_\bq^2+\xi^2)}.
\end{split}
\en
For both of these expressions we repeat the same steps, which lead us to (\ref{sigmap2qr}) with the following results:
\beg\label{sigma34fin}
\delta{\sigma}_3^{\textrm{AL}}(T\ll\Delta_{\textrm{QCP}})\approx\frac{e^2}{2\pi^2}, \quad \delta{\sigma}_4^{\textrm{AL}}(T\ll\Delta_{\textrm{QCP}})\approx-\frac{5e^2}{18\pi^2}.
\en
Next, we proceed with the asymptotic expressions in the thermal regime. 

\paragraph{Thermal regime.} When $T\gg\Delta_{\textrm{QCP}}$ we can make an additional approximation for the fluctuation propagator by expanding the logarithm up to the linear order in $|\varGamma-\varGamma_c|$. Using the notations of Eqs. (\ref{sigma1pp2}) we have
\beg\label{Approx4PropClassic}
\ln\left(\frac{x_\bq-i\xi}{x_c}\right)\approx\frac{\varGamma-\varGamma_c+y_\bq}{\varGamma_c}-\frac{i\xi}{x_c}.
\en 
Then, the remaining integral over $\xi$ can be carried out by using the series representation for the $\sinh^2(\pi\xi)$. It is also easy to verify that in that expansion it will suffice to retain the first-order term only, i.e. we replace $\sin(\pi\xi)\approx\pi\xi$ since the main contribution to the corresponding integral comes from the region of small $\xi$. Then, the integration over $\xi$ can be easily carried out with the following result:
\beg\label{sigma1Classic}
\delta\sigma_1^{\textrm{AL}}(T\gg\Delta_{\textrm{QCP}})\approx\frac{e^2}{4\pi}T\int\limits_0^\infty\frac{y_\bq dy_\bq}{(\varGamma+y_\bq)^2\left(|\varGamma-\varGamma_c|+y_\bq\right)^3}.
\en
Clearly, this integral is divergent at the QCP and so its value is accumulated at small $y_\bq$. One obtains: 
\beg\label{ds1pa}
\delta\sigma_1^{\textrm{AL}}(T\gg\Delta_{\textrm{QCP}})\approx\frac{e^2}{8\pi}\frac{T}{\Delta_{\textrm{QCP}}}.
\en
Thus, we find that the ``classical" AL correction to conductivity in the regime dominated by thermal fluctuations is in fact divergent as one approaches the SC QCP.

We proceed with the calculation of the second contribution in (\ref{sigma1pp2}), which, as we have already mentioned above, accounts for the
contribution from the quantum critical fluctuations as it remains finite in the zero-temperature limit. We note that in the calculation of the term which contains the derivative of the $[L^R]^2$ we only need to retain the leading in $T/T_{c0}$ contribution. Thus, we make the following approximation:
\beg\label{approxderLR2}
\begin{split}
&\textrm{Re}\left\{\frac{i\xi(i\xi-2x_\bq)}{(x_\bq^2+\xi^2)^2}\frac{\partial}{\partial \xi}\left(\left[L^{R}(\bq,-i\xi)\right]^2\right)\right\}\approx
{\nu_F^{-2}}x_c^2\frac{4x_\bq\tilde{x}_\bq \xi}{(\tilde{x}_\bq^2+\xi^2)^2(x_\bq^2+\xi^2)^2}.
\end{split}
\en
The rest of the calculation is similar to the one above: the integral over $\xi$ can be carried out since its value is accumulated in the region where $\xi\coth(\pi\xi)\sim 1$ and the subsequent integration over $y_\bq$ can be carried out exactly as well.  The final result is
\beg\label{ds1ppa} 
\delta\sigma_2^{\textrm{AL}}(T\gg\Delta_{\textrm{QCP}})\approx -\frac{3e^2}{8\pi}\left(12\ln2-\frac{49}{6}\right)\left(\frac{T}{\varGamma_c}\right).
\en
Clearly, this contribution is much smaller as compared to the first one in the thermal regime of temperatures.

The asymptotic expressions for the remaining contributions can be found using exactly the same approximations as the ones we used in the derivation of the results above,
\beg\label{ds2ppa}
\begin{split}
&\delta\sigma_3^{\textrm{AL}}(T)\approx e^2\frac{T}{\varGamma_c}\ln\frac{\varGamma_c}{\Delta_{\textrm{QCP}}}, \quad
\delta\sigma_4^{\textrm{AL}}(T)\approx -\frac{e^2}{3\pi}\frac{T}{\varGamma_c}.
\end{split}
\en
Note that each of the four contributions differs by a type of singular behavior in the limit when $\Delta_{\textrm{QCP}}\to0$.
Thus, we may conclude that in the limit $T\to 0$ the leading Aslamazov-Larkin fluctuation correction to conductivity $\delta\sigma_{2}^{\textrm{AL}}$ increases linearly with temperature.

\section{Asymptotic expressions for $\delta\sigma^{\text{MT}}$}\label{App-MT}

\paragraph{Asymptotic expressions in the quantum regime.} We first consider the limit of low temperatures when $T\ll\Delta_{\textrm{QCP}}$. Employing relations (\ref{GenExpand}) and using the series expansion for the $\coth(\pi\xi)$ [see Eq. (\ref{cothseries}) below] for the regular conductivity correction $\delta\sigma_{\textrm{reg}}^{\textrm{MT}}$ we find
\beg\label{MTregasym}
\begin{split}
&\delta\sigma_{\textrm{reg}}^{\textrm{MT}}(T)\approx\frac{e^2}{4\pi^2}\int\limits_0^\infty dx
\int\limits_0^\infty\frac{\xi d\xi}{2\pi i}\sum\limits_{n=-\infty}^\infty\frac{1}{(n^2+\xi^2)(\xi+i x_\bq)^2\ln[(\varGamma+{\cal D}\bq^2)/\varGamma_c]}.
\end{split}
\en
Here we have neglected the dependence of the fluctuation propagator on $\xi$ since $2\pi T\xi\ll\Gamma_c$.
Integral over $\xi$ can now be performed exactly. Then performing the summation over $n$ and using the fact that $n=0$ term is much smaller than one, we can use (\ref{GenExpand}) to write down the following expression:
\beg\label{MTregasymfin}
\delta\sigma_{\textrm{reg}}^{\textrm{MT}}(T\ll\Delta_{\textrm{QCP}})\approx-\frac{e^2}{8\pi^2}\int\limits_0^\infty\frac{d\eta}{(1+\eta)\ln(1+\eta)}\approx-\frac{e^2}{8\pi^2}\ln\ln\frac{1}{T_{c0}\tau_\pi}.
\en

We now turn our attention to the calculation of the asymptotic expression for the anomalous Maki-Thompson correction (\ref{sigmaMTanom}). Here we again may replace the digamma functions with their asymptotic expression for the large argument, which yields
\beg\label{MTanom1}
\delta\sigma_{\textrm{anom}}^{\textrm{MT}}(T)\approx\frac{{\cal D}e^2}{2\pi}\int\limits_0^\infty
\frac{qdq}{(\frac{\varGamma}{2\pi T}+{\cal D}q^2)}\int\limits_{-\infty}^\infty\frac{d\xi}{\sinh^2(\pi\xi)}\frac{\phi_{\xi}^2}{\left[\ln^2\left(\frac{x_\bq^2+\xi^2}{x_c}\right)+\phi_\xi^2\right]}.
\en
Here we introduced $\tan\phi_\xi=\xi/x_\bq$. Since both $x_c$ and $x_\bq$ are much larger than unity, we may ignore the dependence of the logarithm in the denominator on $\xi$,
\beg\label{approxLog}
\ln\left(\frac{\sqrt{x_\bq^2+\xi^2}}{x_c}\right)\approx\ln\left(\frac{x_\bq}{x_c}\right)
\en
and also approximate $\phi_\xi\approx \xi/x_\bq$.
On account of these approximations, we can perform an integration over $\xi$ using
\beg\label{FamousInt}
\int\limits_{-\infty}^\infty\frac{dx}{\sinh^2x}\frac{x^2}{(x^2+\pi^2r^2)}=2r\psi'(r)-\frac{1}{r}-2.
\en
Since in our case $r=x_\bq\ln(x_\bq/x_c)\gg1$ we may expand the digamma functions in powers of $1/r$,
\beg\label{expanddi}
\psi'(r\gg1)\approx\frac{1}{r}+\frac{1}{2r^2}+\frac{1}{6r^3}.
\en
Inserting this expressions into (\ref{MTanom1}) and rescaling the remaining integration variable yields
\beg\label{MTanom2}
\delta\sigma_{\textrm{anom}}^{\textrm{MT}}(T)\approx
\frac{2e^2}{3}\frac{T^2}{\varGamma_c^2}\int\limits_0^\infty\frac{dy}{\left(\frac{\varGamma}{\varGamma_c}+y\right)^3\ln^2(\frac{\varGamma}{\varGamma_c}+y)},
\en
while the remaining integral is clearly divergent at the critical point $\varGamma=\varGamma_c$ as $1/|\varGamma-\varGamma_c|$. Thus, we conclude that the correction to conductivity (\ref{MTanom2}) is 
\beg\label{MTanom2p}
\delta\sigma_{\textrm{anom}}^{\textrm{MT}}(T\ll\Delta_{\textrm{QCP}})\approx\frac{\pi e^2}{3}\frac{T^2}{\varGamma_c\Delta_{\textrm{QCP}}}.
\en
Thus, we see that the regular contribution provides a leading Maki-Thompson correction at low temperatures. 

\paragraph{Asymptotic expressions in the thermal regime.} Let us now compute the asymptotic Maki-Thompson corrections in the regime dominated by thermal fluctuations above the quantum critical point, $T\gg\Delta_{\textrm{QCP}}$. To determine the anomalous correction to the regular Maki-Thompson contribution we may employ previous approximations. The subsequent integration over $\xi$ yields
\beg\label{MTregT1}
\delta\sigma_{\textrm{reg}}^{\textrm{MT}}(T\gg\Delta_{\textrm{QCP}})\approx-\frac{e^2}{\pi\varGamma_c}\sum\limits_{n=-\infty}^{\infty}\int\limits_0^\infty\frac{{\cal D}qdq}{(|n|+x_\bq)^2\left(|n|+x_\bq-\frac{\varGamma_c}{2\pi T}\right)}.
\en
It is straightforward to verify that the leading contribution to the sum comes from the $n=0$ term. Changing the integration variable to $\eta={\cal D}q^2/\varGamma_c$ and performing an integration over $\eta$ gives
\beg\label{MTregT2}
\delta\sigma_{\textrm{reg}}^{\textrm{MT}}(T\gg\Delta_{\textrm{QCP}})\approx-{2e^2}\left(\frac{T}{\varGamma_c}\right)\ln\left(\frac{\varGamma_c}{\Delta_{\textrm{QCP}}}\right).
\en

The asymptotic expression for the anomalous part of the Maki-Thompson contribution can be found similarly to the one which lead to the expression (\ref{MTanom2}). We can make an additional approximation here for the fluctuation propagator by expanding the logarithm around the quantum critical point
\beg\label{LRexpand}
\ln\left(\frac{\varGamma+{\cal D}q^2-2\pi i T\xi}{\varGamma_c}\right)\approx\frac{\varGamma-\varGamma_c+{\cal D}q^2-2\pi iT\xi}{\varGamma_c}.
\en
Then the expression for $\delta\sigma_{\textrm{anom}}^{\textrm{MT}}$ acquires the following form
\beg\label{MTanomhighT0}
\delta\sigma_{\textrm{anom}}^{\textrm{MT}}(T\gg\Delta_{\textrm{QCP}})\approx\frac{e^2\varGamma_c}{2\pi}\int\limits_0^\infty\frac{{\cal D}qdq}{(\varGamma+{\cal D}q^2)^2}\int\limits_{-\infty}^\infty\frac{\xi^2d\xi}{\sinh^2(\pi\xi)(\tilde{x}_\bq^2+\xi^2)},
\en
where $\tilde{x}_\bq=(\varGamma-\varGamma_c+{\cal D}\bq^2)/2\pi T$. We can now evaluate the integral over $\xi$ using formula (\ref{FamousInt}). In this case, however, the resulting digamma functions is a function of $\tilde{x}_\bq$, which is small in the thermal regime, so we can approximate
\beg\label{approxpsiT}
\psi'(\tilde{x}_\bq\ll1)\approx\frac{1}{\tilde{x}_\bq^2}.
\en
Thus, for the conductivity we may write 
\beg\label{MTanomT1}
\delta\sigma_{\textrm{anom}}^{\textrm{MT}}(T\gg\Delta_{\textrm{QCP}})\approx\frac{e^2}{2\pi}\left(\frac{T}{\varGamma_c}\right)\int\limits_0^\infty\frac{dy}{(1+y)^2\left(\frac{\Delta_{\textrm{QCP}}}{\varGamma_c}+y\right)}\approx\frac{e^2}{2\pi}\frac{T}{\varGamma_c}\ln\frac{\varGamma_c}{\Delta_{\textrm{QCP}}}.
\en
Thus, we find that the thermal fluctuations correction has the linear temperature dependence  compared to the quadratic $T$ dependence found in the regime of the quantum fluctuations (\ref{MTanom2p}). 

\section{Asymptotic expressions for $\delta\sigma^{\text{DOS}}$}\label{App-DOS}

\paragraph{Asymptotic expressions in the quantum regime.} As we have noted above, the first correction to conductivity $\delta\sigma_{1}^{\textrm{DOS}}(T)$ matches the corresponding ``regular" Maki-Thompson correction, so we can readily write down
\beg\label{DOS1symfin}
\delta\sigma_{1}^{\textrm{DOS}}(T\ll\Delta_{\textrm{QCP}})\approx-\frac{e^2}{8\pi^2}\ln\ln\frac{1}{T_{c0}\tau_\pi}.
\en
As it already can be seen by comparing the form of Eq. (\ref{sigma2DOS}) with the corresponding expression for $\delta\sigma_{\textrm{anom}}^{\textrm{MT}}$ the calculation of an asymptotic form for $\delta\sigma_{2}^{\textrm{DOS}}$ is quite similar to the one performed above for 
$\delta\sigma_{\textrm{anom}}^{\textrm{MT}}$. At the intermediate step we find the following expression:
\beg\label{DOS2symfin}
\delta\sigma_{2}^{\textrm{DOS}}(T\ll\Delta_{\textrm{QCP}})\approx-\frac{2e^2}{3}\frac{T^2}{\varGamma_c^2}\int\limits_0^\infty\frac{d\eta}{(1+\eta)^3\ln^2\left(\frac{\varGamma}{\varGamma_c}+\eta\right)}.
\en
Although this integral is clearly divergent at the QCP, the correction itself 
\beg\label{DOS2quantum}
\delta\sigma_{2}^{\textrm{DOS}}(T\ll\Delta_{\textrm{QCP}})\approx-\frac{2e^2}{3}\frac{T^2}{\varGamma_c\Delta_{\textrm{QCP}}}
\en
is much smaller compared to (\ref{DOS1symfin}).

\paragraph{Asymptotic expressions in the thermal regime.} In this regime the first DOS correction is the same as the regular Maki-Thompson correction,
\beg\label{sigmaDOS1highT}
\delta\sigma_{1}^{\textrm{DOS}}(T\gg\Delta_{\textrm{QCP}})\approx{2e^2}\left(\frac{T}{\varGamma_c}\right)\ln\left(\frac{\varGamma_c}{\Delta_{\textrm{QCP}}}\right).
\en
Consequently, for the second DOS correction the following expression can be found:
\beg\label{sigmaDOS2highT}
\delta\sigma_{2}^{\textrm{DOS}}(T\gg\Delta_{\textrm{QCP}})\approx-\frac{{\cal D}e^2}{\pi T}\left(\frac{\varGamma_c}{2\pi T}\right)\int\limits_0^\infty\frac{qdq}{2\pi}\int\limits_{-\infty}^\infty\frac{\xi^2d\xi}{\sinh^2(\pi\xi)(x_\bq^2+\xi^2)(\tilde{x}_\bq^2+\xi^2).}
\en
Here $\tilde{x}_\bq=(\varGamma-\varGamma_c+{\cal D}\bq^2)/2\pi T$. Since the leading contribution to this integral comes from the region of small $\xi$, in the thermal regime we can evaluate the integral over $\xi$ approximately by expanding $\sinh(\pi\xi)\approx\pi \xi$. It then follows
\beg\label{DOS2T}
\delta\sigma_{2}^{\textrm{DOS}}(T\gg\Delta_{\textrm{QCP}})\approx-\left(\frac{e^2}{\pi}\right)\frac{T}{\varGamma_c}\ln\frac{\varGamma_c}{\Delta_{\textrm{QCP}}}.
\en
This result suggests that this correction dominates the anomalous Maki-Thompson correction in the thermal fluctuations dominated regime. 


\section{Diffusion coefficient renormalization: $\delta\sigma^{\textrm{DCR}}$}\label{App-DCR12}
In this section we derive the asymptotic expressions for the corrections to dc-conductivity schematically represented by the diagrams in Fig. \ref{FigDCR}. As we have already discussed in the main text, the contribution from the third and fourth diagram is given by the same expressions as the first two. Thus, we start with the analysis of the contribution from the first two diagrams. 
\paragraph{Quantum fluctuations regime.} We start with the analysis of $\delta\sigma_\pi^{\textrm{DCR}_1}$, 
\beg\label{dsigmaDCRpiC}
\delta\sigma_{1,\pi}^{\textrm{DCR}}=\frac{2\sigma_{D}}{(4\pi)^3T^2}\int\limits_{-\infty}^\infty\frac{d\xi}{2i}\coth(\pi \xi)\int\frac{{\cal D}q_x^2d^2\bq}{(2\pi)^2}L^R(\bq,- i\xi)\psi'''\left(\frac{1-i\xi+x_\bq}{2}\right),
\en
where $L^R(\bq,-i\xi)$ is the retarded part of the fluctuation propagator and function $x_\bq=\varGamma_\bq/2\pi T$ is introduced for convenience. In this regime the logarithmic digamma function and the expression for the fluctuation propagator can be approximated as follows:
\beg\label{ApproxQCR}
\psi'''\left(\frac{1-i\xi+x_\bq}{2}\right)\approx\frac{8}{(x_\bq-i\xi)^3}, \quad 
L^R(\bq,-i\xi)=\nu_F^{-1}\ln^{-1}\left(\frac{x_\bq-i\xi}{\alpha_c}\right).
\en
Here we introduced a parameter $\alpha_c=\Gamma_\pi^{(c)}/2\pi T$ for brevity. 
The analysis is greatly simplified if we represent $\coth(\pi \xi)$ as 
\beg\label{cothseries}
\coth(\pi\xi)=\frac{1}{\pi}\sum\limits_{n=-\infty}^\infty\frac{\xi}{n^2+\xi^2}.
\en
At low temperatures ($T\ll \Delta_{\textrm{QCP}}$), in the expression for the fluctuation propagator we can ignore the dependence of the logarithm on $\xi$. Then, inserting (\ref{cothseries}) into (\ref{dsigmaDCRpiC}), using approximations (\ref{ApproxQCR}) along with the expressions above and performing an integration over $\xi$ yields
\beg\label{dsigmaDCRpiC2}
\delta\sigma_{1,\pi}^{\textrm{DCR}}(T\ll \Delta_{\textrm{QCP}})\approx\frac{e^2}{(4\pi)^2}\sum\limits_{n=-\infty}^\infty\int\limits_0^\infty\frac{ydy}{(|n|+\alpha+y)^3}\ln^{-1}\left(\frac{\alpha+y}{\alpha_c}\right).
\en
Since $n=0$ zero term gives negligibly small contribution $\sim O(T/\Gamma_\pi^{(c)})$, we can change the summation over $n\geq 0$, which to the leading order yields
\beg\label{dsigmaDCRpiC3}
\delta\sigma_{1,\pi}^{\textrm{DCR}}(T\ll\Delta_{\textrm{QCP}})\approx\frac{e^2}{(4\pi)^2}\int\limits_0^\infty\frac{ydy}{(\alpha+y)^2}\ln^{-1}\left(\frac{\alpha+y}{\alpha_c}\right)\approx\frac{e^2}{(4\pi)^2}\ln\ln\left(\frac{1}{T_{\textrm{c}0}\tau_{\textrm{t}}}\right).
\en
Note that this correction, unlike the corresponding corrections from $\textrm{MT}$ and $\textrm{DOS}$ diagrams, is actually positive. 

Similar analysis of the second term $\delta\sigma_{1,0}^{\textrm{DCR}}(T)$ can be only performed numerically, since the argument of the digamma function $\psi(1/2+{\cal D}\bq^2/2\pi T-i\xi)$ is of the order $O(1)$ at long wavelengths and, therefore, we can not use the asymptotic expansion. However, it is straightforward to compute both $\delta\sigma_{1,0;1,\pi}^{\textrm{DCR}}(T)$ for any temperature range and we found that its contribution is much smaller than that of $\delta\sigma_{1,\pi}^{\textrm{DCR}}(T)$.

\paragraph{Thermal fluctuations regime.} In the regime dominated by thermal fluctuations, we can further approximate the expression for the pair fluctuation propagator by expanding it around $\alpha=\alpha_c$ and retaining the first-order term only,
\beg\label{approxLT}
L^R(\bq,-i\xi)\approx\nu_F^{-1}\frac{\alpha_c}{\alpha-\alpha_c+\frac{{\cal D}\bq^2}{2\pi T}-i\xi}.
\en
We can now repeat the same steps of the ones, which lead to the expression (\ref{dsigmaDCRpiC2}). Given (\ref{approxLT}) 
we again employ the expansion (\ref{cothseries}) and integrate over $\xi$ which yields
\beg\label{dsigmaDCRpiT1}
\delta\sigma_{1,\pi}^{\textrm{DCR}}(T\gg\Delta_{\textrm{QCP}})\approx\frac{e^2\alpha_c}{(4\pi)^2}\sum\limits_{n=-\infty}^\infty\int\limits_0^\infty\frac{ydy}{(|n|+\alpha+y)^3\left(|n|+\alpha-\alpha_c+y\right)}.
\en
The $n=0$ term in this expansion is equivalent to approximating $\coth(\pi\xi)$ with $(\pi\xi)^{-1}$, which is relevant for the temperature regime that we consider. This leads to 
\beg\label{sMTDCTn0}
\left[\delta\sigma_{1,\pi}^{\textrm{DCR}}(T\gg\Delta_{\textrm{QCP}})\right]_{n=0}\approx\frac{e^2}{32\pi^2}\left(\frac{T}{\varGamma_c}\right).
\en
The calculation of the contribution from $n\not =0$ is done similarly to the one above: we add/subtract to/from the sum in (\ref{dsigmaDCRpiT1}) the $n=0$ term. Then the summation over $n$ can be trivially performed. The integration yields exactly the same expression as (\ref{sMTDCTn0}), but with the opposite sign, so that for the overall correction we find
\beg\label{sMTDCRfin}
\delta\sigma_{1,\pi}^{\textrm{DCR}}(T\gg\Delta_{\textrm{QCP}})\approx\frac{e^2}{16\pi^2}\left(\frac{T}{\varGamma_c}\right).
\en
Notably, all the DCR contributions are positive. 

\bibliography{bibscqcp}

\end{document}